\documentclass[3p, sort&compress, times]{elsarticle}
\usepackage{hyperref}
\usepackage{xspace}
\usepackage{siunitx}
\usepackage{lineno}
\usepackage{mhchem}
\usepackage{bm}

\let\originalleft\left
\let\originalright\right
\renewcommand{\left}{\mathopen{}\mathclose\bgroup\originalleft}
\renewcommand{\right}{\aftergroup\egroup\originalright}

\DeclareMathOperator{\sgn}{sgn}

\hypersetup{colorlinks=true, linkcolor=blue, citecolor=blue, urlcolor=blue}

\providecommand\ito{\^{I}to\xspace}
\providecommand\e{\text{e}}
\providecommand\ppc{\text{ppc}}
\providecommand\M{\text{M}}
\providecommand\dt{\ensuremath{\textrm{d}t}}
\renewcommand\d{\ensuremath{\textrm{d}}}

\begin{document}

\title{Stochastic and self-consistent 3D modeling of streamer discharge trees with Kinetic Monte Carlo}
\author{Robert Marskar}
\address{SINTEF Energy Research, Sem S\ae lands vei 11, 7034 Trondheim, Norway.}
\ead{robert.marskar@sintef.no}
\date{\today}

\begin{abstract}
  This paper contains the foundation for a new Particle-In-Cell model for gas discharges, based on \ito diffusion and Kinetic Monte Carlo (KMC).
  In the new model the electrons are described with a microscopic drift-diffusion model rather than a macroscopic one.
  We discuss the connection of the \ito-KMC model to the equations of fluctuating hydrodynamics and the advection-diffusion-reaction equation which is conventionally used for simulating streamer discharges.
  The new model is coupled to a particle description of photoionization, providing a non-kinetic all-particle method with several attractive properties, such as:
  1) Taking the same input as a fluid model, e.g. mobility coefficients, diffusion coefficients, and reaction rates. 
  2) Guaranteed non-negative densities.
  3) Intrinsic support for reactive and diffusive fluctuations. 
  4) Exceptional stability properties.
  The model is implemented as a particle-mesh model on cut-cell grids with Cartesian adaptive mesh refinement.
  Positive streamer discharges in atmospheric air are considered as the primary application example, and we demonstrate that we can self-consistently simulate large discharge trees.

\end{abstract}

\maketitle


\section{Introduction}
\label{sec:Introduction}
Substantial efforts have been made in order to understand the nature of streamer discharges \cite{Nijdam2020}, which is a specific type of transient and filamentary plasma.
Streamers occur naturally as precursors to electric sparks, and also appear as sprite discharges in the upper atmosphere \cite{P.1998, Stenbae-Nielsen2000, Marshall2006, Ebert2010}.
They are also highly useful for $\textrm{CO}_2$ conversion \cite{VanLaer2015, Zhang2018,Wang2018,doi:10.1002/ente.201500127}, in plasma assisted combustion \cite{Starikovskaia2006, Adamovich2009, Aleksandrov2009, Breden2013, Starikovskaia2014, Starikovskiy2015}, plasma catalysis \cite{1167639, 18870, 4504897, 55956, Nair2004, Grymonpre2001}, and plasma medicine \cite{Laroussi2018}.
Streamers are inherently three-dimensional structures that usually appear in bundles or in the shape of discharge trees.
These develop through repetitive branching, which is a fundamental property of streamers \cite{Ebert2002, Arrayas2002}.
For positive streamer discharges in air, the amount of photoionization in front of the streamer strongly affects the degree of branching \cite{Bagheri_2019, Marskar2020}.
When more photoelectrons are generated in front of positive streamers the amount of streamer branching is reduced.

\begin{figure}[h!t!b!]
  \centering
  \includegraphics{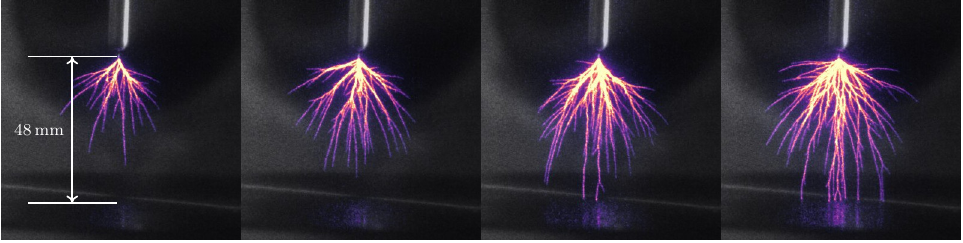}
  \caption{
    Example images of positive streamer discharge trees in technical air at 1 bar pressure.
    The images show the accumulated UV and optical light emission from the discharge (each frame shows a different discharge).
  }
  \label{fig:Nijdam}
\end{figure}

Although single streamers are now relatively well understood, discharge trees are the more relevant structures since they appear in virtually all applications involving streamer discharges.
Figure~\ref{fig:Nijdam} shows an example of such a structure for a positive streamer discharge in air at atmospheric pressure and temperature. 
Clearly, this structure requires full 3D modeling over many orders of magnitude in both space and time, and is therefore quite difficult to describe quantitatively.
Most contemporary computational models can only solve for at most a few filaments.
See e.g. \cite{Marskar2019, Marskar2019b, Marskar2020, Teunissen2017, Bagheri_2019, Lin2020} for recent results with fluid models, or \cite{Teunissen2016, Fierro2016, Fierro2018, Stephens2018, Kohn2018a} for kinetic particle models.

Several researchers have questioned the feasibility of using fluid and particle models for obtaining numerical solutions that describe entire discharge trees \cite{Nijdam2020, Pavan2020, Vazquez2021}, such as those in figure~\ref{fig:Nijdam}.
Given the difficulties in simulating even just an isolated streamer filament \cite{Bagheri2018}, it is easy to understand why such claims are made.
In a recent review \citet{Nijdam2020} remarked that although fluid and particle simulations of streamers with tens or hundreds of branches are computationally unfeasible, reduced-order models of single filaments \cite{Luque2017, Pavan2020} are candidates for improved tree and fractal based models \cite{Niemeyer1984, Akyuz2003, Luque2014, Gonzalez2019}.
Although these models have been heralded for quite some time and could be used for simulating discharge trees, they are still in their infancy and they are unfortunately also excessively simplified.
Many natural phenomena like branching and charge transport do not self-consistently evolve from the model itself, and the lack of a density (or particle) description also complicates quantitative descriptions of the chemistry in the streamer channels.

Currently, drift-diffusion fluid models in the local field approximation (LFA) are most frequently used for studying streamer discharges.
With fluid models the advection-diffusion-reaction equation for the electron density is discretized on an Eulerian grid, and the plasma density is updated in time using either explicit or implicit time integration \cite{Bagheri2018}.
There are several well-known numerical restrictions for fluid models, such as the existence of a Courant-Friedrichs-Lewy (CFL) condition on the time step $\Delta t$, or restriction by the dielectric relaxation time $\epsilon_0/\sigma$ where $\sigma$ is the conductivity of the plasma.
The latter can be avoided by using semi-implicit formulations \cite{Ventzek1994}.
Infrequently mentioned is the fact that there is a rather fundamental requirement on the spatial resolution $\Delta x$ as well \cite{Villa2014}, which applies to both explicit and implicit temporal discretizations.
One issue that is often faced in simulation codes is that explicit codes at best have time steps $\Delta t \propto \Delta x$, which leads to an undesired scaling of computational resources.
For example, refining the grid $\Delta x \rightarrow \Delta x/2$ doubles the amount of grid cells per coordinate direction, and requires twice as many time steps.
Implicit codes can decouple $\Delta t$ from $\Delta x$, but it is not clear how to obtain a scalable implicit discretization in the context of the frequent regridding that is a de-facto requirement for large scale 3D streamer discharge simulations \cite{Teunissen2017,Marskar2019, Marskar2019b, Marskar2020}.

Cognizant of the above issues, we have developed a new model based on a \emph{microscopic} drift-diffusion model rather than a macroscopic one.
This is combined with mesoscopic reaction algorithms for describing the stochastic plasma chemistry, i.e. we replace the conventionally used deterministic chemistry by a Kinetic Monte Carlo (KMC) algorithm. 
The new model takes particle discreteness, random collisions, and stochastic reactions into account.
Fundamentally, the model is a non-kinetic Particle-In-Cell (PIC) model, and it is indeed ironic that this is actually a helpful model since the switch from a fluid to a particle description is usually associated with an increased computational cost.
But the new model has no fundamental restriction on $\Delta x$ or $\Delta t$, and numerical tests show that it is exceptionally stable in both space and time.
Importantly, we are also achieving a decoupling of $\Delta t$ from $\Delta x$ without requiring an implicit discretization.
This renders the new model capable of obtaining numerical solutions not only for single filaments, but also for comparatively large discharge trees.
In this paper we use the model to investigate laboratory discharges, but the model itself is applicable to many other types of streamers (e.g. sprites).

This paper has two main goals:
1) A thorough presentation of the model, with details as to how it can be implemented with robust and scalable computer algorithms.
2) A capability demonstration for self-consistent simulation of discharge trees at the laboratory scale, similar to the ones shown in figure~\ref{fig:Nijdam}.
The organization of this paper is as follows.
Section~\ref{sec:prelude} presents a computational prelude that focuses on finite-volume discretization issues for fluid models, for both explicit and implicit time discretizations.
In section~\ref{sec:model} we formulate the new model and discuss its connection to the conventional fluid model.
Section~\ref{sec:implementation} contains the numerical discretization of the model.
In section~\ref{sec:examples} we provide some numerical tests of the model, and some concluding remarks are provided in section~\ref{sec:conclusion}.

\section{Prelude}
\label{sec:prelude}

We first consider an underlying issue facing the stability properties of discretized fluid models in the LFA.
Our line of reasoning follows \citet{Villa2014} who proved that the spatial resolution for fluid models must essentially resolve the avalanche length for the solution to remain bounded in time.
Consider a one-dimensional advection-reaction model for the electron density $n$, for the moment ignoring electron diffusion:

\begin{equation}
  \label{eq:advection1D}
  \partial_t n = -v\partial_x n + \alpha v n,
\end{equation}
where $v > 0$ is a constant electron velocity and $\alpha > 0$ is a constant ionization coefficient.
The exact solution is

\begin{equation}
  \label{eq:advection_reaction_exact}
  n(x,t) = \widetilde{n}(x-vt)\exp\left(\alpha x\right),
\end{equation}
where $\widetilde{n}(x-vt)$ is some initial function.
If $\widetilde{n}$ is bounded in time, so is $n(x,t)$.

We now consider the numerical discretization of equation~\eqref{eq:advection1D} on a one-dimensional Cartesian grid with grid points $x_i = i\Delta x$ where $\Delta x$ is the grid point spacing and $i$ is the grid index.
Each grid cell spans the volume $[x_i-\Delta x/2,x_i+\Delta x/2]$.
A first order finite-volume upwind discretization in space with an implicit Euler discretization in time for equation~\eqref{eq:advection1D} yields

\begin{equation}
  \label{eq:fluid_implicit}
  \begin{split}
    n_{i}^{k+1} &= \frac{n_{i}^k}{1 + \frac{v\Delta t}{\Delta x} - \alpha v\Delta t} + \frac{v\Delta t}{\Delta x}n_{i-1}^{k+1}\\
    &\geq \frac{n_{i}^k}{1 + \xi\left(1 - \alpha \Delta x\right)},
  \end{split}
\end{equation}
where $\xi\equiv v\Delta t/\Delta x \geq 0$ is the Courant number.
The solution $n_{i}^{k+1}$ is bounded in time only if

\begin{equation}
  \alpha \Delta x \leq 1.
\end{equation}
Here, the discretization is fully implicit but it is only conditionally stable and non-negative.

More generally, the underlying stability issues are related to the advective-reactive coupling.
Using a Godunov splitting for equation~\eqref{eq:advection1D} with explicit fractional Euler steps yields

\begin{align}
  \label{eq:fluid_godunov}  
  n_{i}^\dagger &= n_{i}^k(1 - \xi) + \xi n_{i-1}^k, \\
  \begin{split}
    n_i^{k+1} & = n_i^\dagger + \alpha v\Delta t n_i^\dagger  \\
    &= \left[n_i^k\left(1-\xi\right) + \xi n_{i-1}^k\right]\left(1+\xi\alpha\Delta x\right)\\
    & \geq n_i^k\left(1-\xi\right)\left(1+\xi\alpha\Delta x\right).
  \end{split}
\end{align}
The stability region is now $\alpha\Delta x \leq 1/(1-\xi)$, i.e. the discretization is more stable for larger time steps.
Splitting methods expand the stability region because they advect electrons out of the grid cell before they react.

Several codes \cite{Bagheri2018,Marskar2019} use second order slope-limited discretizations, but these discretization are not fundamentally more stable.
An analysis is more difficult in this case, but one only needs to observe that slope-limited schemes default to piecewise constant reconstruction if there is a local maximum or a large gradient in the solution, and in this case one again obtains the stability limit $\alpha \Delta x \leq 1$.
The problem dimensionality and presence of diffusion also affects the stability region.
However, since equation~\eqref{eq:advection1D} is a subset of multi-dimensional simulations where $v_y=v_z=0$ and $v_x \neq 0$, the stability limit applies to multi-dimensional simulations as well.

The analysis above is quite simplified and ignores the fact that the streamer is a moving structure and thus that any potential instability regions $\alpha \Delta x > 1$ move with the solution.
In practice, the situation is far less dire and one may still observe that $n_i^k$ remains bounded even for quite significant violations of $\alpha \Delta x \leq 1$.
It is nonetheless clear that caution is needed for fluid simulations since numerical underresolution is fundamentally capable of enabling unbounded growth in the plasma density and corresponding non-physical diverging growth of the electric field $E$, i.e. $E\rightarrow \infty$ as $n\rightarrow \infty$.
Such instabilities have been scrutinized in recent years, and they are particularly relevant in the cathode sheath \cite{Niknezhad2021} and in the context of so-called stagnant positive streamers \cite{Pancheshnyi2004}.
Although the LFA is often identified as the culprit \cite{Niknezhad2021}, the underlying issue is also present in the transport equation itself.
Furthermore, the ionization coefficient $\alpha$ increases with $E$, and in practice the requirement on $\Delta x$ introspectively depends on the numerical solution itself, complicating the selection of a spatial step size.

Various resolutions to the spatial stability restriction have been proposed in the literature.
\citet{Villa2014} showed that the mesh-dependent stability criterion $\alpha\Delta x \leq 1$ can be removed by treating reactions with an upwind method.
\citet{Marskar2020} softened it by using a Godunov splitting like equation~\eqref{eq:fluid_godunov}, and used a Corner Transport Upwind (CTU) scheme \cite{Colella1990} for maintaining an overall larger time step for multi-dimensional simulations.
Model corrections to the ionization term have also been considered.
\citet{Niknezhad2021} change the characteristic length scale of the ionization term by applying a smoothing operator that changes $\alpha$ without net loss in the number of reactions.
\citet{Soloviev2014, Marskar2020,Teunissen2020}, and \citet{Li2022} have considered alterations to $\alpha$ based on electron energy considerations.

\begin{figure}[h!t!b!]
  \centering
  \includegraphics{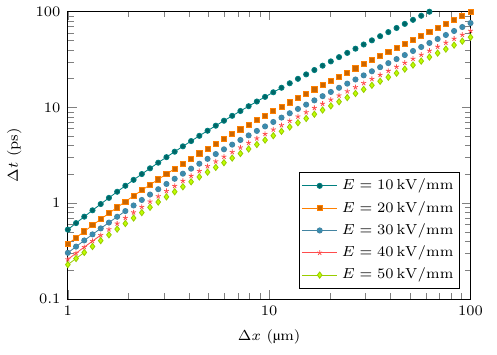}
  \caption{Computed time step for 3D fluid models using equation~\eqref{eq:explicitDt} for discharges in atmospheric air.
    The combined line-mark style indicates the region $\alpha \Delta x \leq 1$ while the marks-only style indicates the region $\alpha\Delta x > 1$.
    Velocities are given by $v_x = v_y = v_z = \mu E/\sqrt{3}$ where $E$ is the electric field magnitude, and the electron mobility $\mu$ and diffusion coefficient $D$ are obtained using BOLSIG+ \cite{Hagelaar2005a} and the SIGLO database \cite{SigloDB}.
    }
  \label{fig:FluidTimeStep}
\end{figure}

For explicit codes, the requirement on $\Delta x$ can be quite penalizing for the time steps that can be used.
On structured Cartesian 3D grids $\Delta x = \Delta y = \Delta z$, a fully explicit discretization using a first order upwind method and centered finite differencing of the diffusion operator yields the time step restriction

\begin{equation}
  \label{eq:explicitDt}
  \Delta t \leq \left(\frac{|v_x| + |v_y| + |v_z|}{\Delta x} + \frac{6D}{\Delta x^2}\right)^{-1},
\end{equation}
where we now also include the electron diffusion coefficient $D$.
Figure~\ref{fig:FluidTimeStep} shows how this time step varies with $\Delta x$ for different selections of the field strength $E$.
\citet{Marskar2020} found that $E\sim\SI{25}{\kilo\volt\per\milli\meter}$ for streamer discharges in air, and with a spatial resolution of $\Delta x \lesssim \SI{2}{\micro\meter}$ the time step is approximately \SI{1}{\pico\second}.
A reasonable simulation time for streamer discharges in atmospheric air is around \SI{100}{\nano\second}, requiring roughly \num{100000} time steps, at which point even fluid simulations become numerically expensive. 

Implicit methods are subject to the same requirement on $\Delta x$ as explicit methods, but they remain attractive since they do not impose fundamental limitations on $\Delta t$.
Unfortunately, 3D simulations often use hundreds of millions of grid points \cite{Marskar2019, Marskar2019b, Marskar2020} and billions of degrees of freedom.
Full Newton methods \cite{Arslanbekov2021} are not very practical at this scale since the full Jacobian must be factored at every time step.
Jacobian-Free Newton-Krylov (JFNK) is a more attractive computational strategy, but it is not clear if JFNK methods remain computationally feasible at this scale, particularly when adaptive mesh refinement (AMR) is required. 

Next, consider the evolution of the microscopic version of equation~\eqref{eq:advection1D}:
\begin{subequations}
  \label{eq:ito_kmc1D}    
  \begin{align}
    \d_t X &= v, \\
    \d_tW &= \alpha v W,
  \end{align}
\end{subequations}
where $X$ is a one-dimensional electron position and $W$ is the (average) number of electrons sharing this position.
For demonstration purposes, we are using a deterministic reaction rate equation for $W$.
This is less meaningful when dealing with a particle method but we improve on this aspect later in the paper. 
Consider a single starting electron, $X(0) = 0$, $W(0) = 1$, in which case the solutions to equation~\eqref{eq:ito_kmc1D} using the explicit Euler rule until time $t = k\Delta t$ are

\begin{align}
  X^k &= v t, \\
  W^k &= \left(1 + \alpha v \Delta t\right)^{k}.
\end{align}
The number of particles per unit length for a grid cell $i$ is then

\begin{equation}
  \label{eq:ito_exact}  
  n_i^k=
  \begin{cases}
    \frac{1}{\Delta x}\left(1 + \alpha v \Delta t\right)^k & \text{if $\left|vt - x_i\right| \leq \Delta x/2$},\\
    0 & \text{otherwise}.
  \end{cases}
\end{equation}
This is to be contrasted with the exact solution to equation~\eqref{eq:advection1D} with the initial condition $\widetilde{n}(x -vt) = \delta(x-vt)$,

\begin{equation}
  \label{eq:advection_exact}
  \begin{split}
    n_i(t) &= \frac{1}{\Delta x_i}\int_{x_i -\Delta x/2}^{x_i + \Delta x_2} \delta(x-vt)\exp\left(\alpha x\right)\d x \\
    &=
    \begin{cases}
      \frac{1}{\Delta x}\exp\left(\alpha v t\right) & |vt - x_i| \leq \Delta x/2, \\
      0 & \text{otherwise}.
    \end{cases}
  \end{split}
\end{equation}
The two solutions (equations~\eqref{eq:ito_exact} and \eqref{eq:advection_exact}) differ only due to the way we approach the numerical integration of equation~\eqref{eq:ito_kmc1D}.
Notably, numerical discretizations that start from equation~\eqref{eq:ito_kmc1D} do not require $\alpha \Delta x \leq 1$, and we can identify why:
Equation~\eqref{eq:fluid_implicit} is numerically diffusive and the electron density only asymptotically tends to zero as $t\rightarrow\infty$, and thus there is always some fraction of $n$ that will react in the grid cell \cite{Villa2014}.
On the other hand, there is no numerical diffusion involved in equation~\eqref{eq:ito_exact} and the discretization is also stable for any time step, i.e. it does not have a CFL condition.
These are the two basic properties that we exploit in the new PIC model.

\section{The new model}
\label{sec:model}

\subsection{Particle transport}

Rather than using a macroscopic drift-diffusion model for the electrons, which is subject to fairly strict requirements on $\Delta x$ and $\Delta t$, we consider a \emph{microscopic} model based on \ito diffusion

\begin{equation}
  \label{eq:ito}
  \d\bm{X}_p = \bm{V}_p\dt + \sqrt{2D_p}\d\bm{W}^p_t,
\end{equation}
where $\bm{X}_p$ is the position of a particle $p$, $\bm{V}_p$ is the drift velocity of the particle and $\sqrt{2D_p}$ is the diffusion coefficient of the particle.
Here, $\d\bm{W}_t^p$ is a Wiener process over a time $\dt$.
It can be represented as $\d\bm{W}_t^p = \sqrt{\dt}\bm{\mathcal{N}}_p$ where $\bm{\mathcal{N}}_p$ is a normal distribution with standard deviation of 0 and variance of 1 in $d$-dimensional physical space. 
The noise is uncorrelated in time and space, and independent of noise acting on other particles. 
The representation of the particle diffusion coefficient as $\sqrt{2D_p}$ is due to a convenient normalization when coarse-graining the model onto a continuum representation where the macroscopic diffusion coefficient $D$ appears instead.

Averaging equation~\eqref{eq:ito} over many identical particles, i.e. $\bm{V}_p = \bm{v}$, $D_p= D$, yields

\begin{subequations}
  \label{eq:ensemble_averages}
  \begin{align}
    \label{eq:advective_average}
    \left\langle\bm{X}_p(t+\Delta t) - \bm{X}_p(t)\right\rangle &= \bm{v}\Delta t, \\
    \label{eq:diffusive_average}
    \left\langle\left[\bm{X}_p(t+\Delta t) - \bm{X}_p(t) - \bm{v}\Delta t\right]^2\right\rangle &= 2Dd\Delta t,
  \end{align}
\end{subequations}
where $\left\langle\ldots\right\rangle$ indicates the expectation value. 
Thus, by taking $\bm{V}$ and $D$ to be the macroscopic electron drift velocity and diffusion coefficients, the \ito model recovers macroscopic drift-diffusion statistics. 
In this paper we adopt the LFA and take $\bm{v}$ and $D$ to be functions of $\bm{E}$. 
The velocity and diffusion coefficients are found by interpolation of the macroscopic quantities $\bm{v}$ and $D$ to the particle positions, i.e.

\begin{align}
  \bm{V}_p &= \bm{v}\left(\bm{X}_p\right), \\
  D_p &= D\left(\bm{X}_p\right).
\end{align}
Extensions to the local mean energy approximation where the coefficients are given as functions of the average electron energy are not examined in this paper. 

In a formal derivation \citet{Dean1996} showed that the evolution of the global density

\begin{equation}
  n(\bm{x},t) = \sum_p\delta\left[\bm{x} -\bm{X}_p(t)\right]
\end{equation}
yields the advection-diffusion equation of fluctuating hydrodynamics: 

\begin{equation}
  \label{eq:sadr}
  \frac{\partial n}{\partial t} = \nabla\cdot\left(-\bm{v}n+ D\nabla n + \sqrt{2Dn}\bm{Z}\right),
\end{equation}
where $\bm{Z}(\bm{x},t)$ is a Gaussian random field without space-time correlations,

\begin{equation}
  \langle \bm{Z}(\bm{x},t)\bm{Z}(\bm{x}^\prime,t^\prime)\rangle = \delta\left(\bm{x}-\bm{x}^\prime\right)\delta\left(t-t^\prime\right).
\end{equation}
The term $\sqrt{2Dn}\bm{Z}$ is a stochastic flux that accounts for fluctuations from Brownian motion. 
This term is usually ignored in studies of non-equilibrium gas discharges.
In the macroscopic limit of vanishing fluctuations, equation~\eqref{eq:sadr} yields the deterministic advection-diffusion equation which is the conventional starting point for fluid models.
The recovery of equation~\eqref{eq:sadr} from equation~\eqref{eq:ito} in the macroscopic limit is hardly surprising since equation~\eqref{eq:ito} is a \emph{microscopic} drift-diffusion model.

\subsection{Kinetic Monte Carlo}

\label{sec:plasma_chemistry}

For the plasma chemistry we compute reactions locally within each grid cell, using KMC.
Suppose that we are provided with a set of reactions that evolve a system of $M$ different chemical species $S_i,\,i\in[1,2,.\ldots,M]$.
The number of particles for each species $S_i$ is given by a state vector

\begin{equation}
  \vec{X}(t) = \begin{pmatrix}
    X_1(t)\\
    X_2(t)\\
    \vdots\\
    X_M(t)
  \end{pmatrix},
\end{equation}
where $X_i(t)$ is the number of particles of type $S_i$ in some computational volume $\Delta V$ at time $t$.
Reactions are represented stoichiometrically, e.g.

\begin{equation}
  \label{eq:reaction_example}
  S_A + S_B + \ldots \xrightarrow{k} S_C + S_D + \ldots,
\end{equation}
where $k$ is the reaction rate. 
The set of such reactions is called the \emph{reaction network} $\vec{R}$.
Let $\vec{\nu}_{r}$ denote the state change in $\vec{X}$ caused by a single firing of a reaction of type $r$, i.e. 

\begin{equation}
  \vec{X}\Rightarrow \vec{X} + \vec{\nu}_r,
\end{equation}
For example, if $\vec{X} = (X_1, X_2)^\intercal$ and the reaction network consists of a single reaction $S_1 \xrightarrow{} S_2$ then $\vec{\nu}_1 = (-1,1)^\intercal$.

Propensity functions $a_r\left(\vec{X}(t),t\right)\dt$ are defined as the probability that exactly one reaction of type $r$ occurs in the infinitesimal interval $[t,t+\dt]$.
In other words, $a_r$ can loosely be interpreted as the number of reactions of type $r$ per unit time in a computational volume.
The rates $k$ that occur in reactions like equation~\eqref{eq:reaction_example} are not equivalent to the conventional reaction rates that are used in the deterministic reaction rate equation (RRE).
For a unipolar reaction of the type $S_1 \xrightarrow{k} \varnothing$ the propensity function is $a_r = kX_1$ and in this case $k$ is numerically equal to the rate that occurs in the RRE (see equation~\eqref{eq:rre}).
However, for bipolar reactions of the type $S_1 + S_1 \xrightarrow{k} \varnothing$ the propensity is $k\frac{1}{2} X_1(X_1-1)$ since there are $\frac{1}{2}X_1(X_1-1)$ unique pairs of particles of type $S_1$.

We use the KMC algorithm as proposed by \citet{Cao2005, Cao2006}. 
This algorithm advances $\vec{X}$ over a time step $\Delta t$ using a sequence of adaptive smaller steps $\Delta \tau$ where the reaction network is advanced using either the SSA \cite{Gillespie1977}, tau-leaping, or a combination of these.
This algorithm is discussed further in section~\ref{sec:reaction_algorithm}, but we first provide some context to the SSA/KMC and tau-leaping algorithms.

\subsubsection{Stochastic simulation algorithm (SSA) and tau-leaping}
The SSA (or Gillespie algorithm \cite{Gillespie1977}), is a next-reaction model which advances $\vec{X}(t)$ one reaction at a time.
Given a total propensity $A(t) = \sum_r a_r\left(\vec{X}(t),t\right)$, the time until the next reaction is randomly determined from

\begin{equation}
  \Delta T_{\text{next}} = \frac{1}{A(t)}\ln\left(\frac{1}{u_1}\right),
\end{equation}
where $u_1\in[0,1]$ is a random number sampled from a uniform distribution.
The reaction type is further determined with

\begin{equation}
  j = \text{smallest integer satisfying} \sum_{r=0}^{j-1} a_r\left(\vec{X}(t),t\right)> u_2 A(t),
\end{equation}
where $u_2\in[0,1]$ is another uniformly sampled random number.
The system is then advanced as

\begin{equation}
  \vec{X}\left(t+\Delta T_{\text{next}}\right) = \vec{X}(t) + \nu_j.
\end{equation}
The SSA resolves one reaction at a time, and the algorithm becomes increasingly inefficient as the number of reactions per unit time grows.
In its isolated form, the algorithm is not very useful for discharge simulations.

The tau-leaping method advances the entire reaction network in a single step over time $\Delta t$ using Poisson sampling:

\begin{equation}
  \label{eq:tau_leaping}
  \vec{X}\left(t+\Delta t\right) = \vec{X}(t) + \sum_r\vec{\nu}_r \mathcal{P}\left[a_{r}\left(\vec{X}(t),t\right)\Delta t\right],
\end{equation}
where $\mathcal{P}\left(\mu\right)$ is a random number sampled from a Poisson distribution with mean $\mu$.
If the propensity functions $a_r$ do not change significantly on the time interval $[t,t+\Delta t]$ then reaction events are statistically independent, which is the condition for the validity of the tau-leaping scheme.
A tau-leaping method has been considered by \citet{Luque2011} in the context of streamer discharges (although the authors do not use the tau-leaping terminology).
Unlike the SSA, equation~\eqref{eq:tau_leaping} does not guarantee a physically valid state since Poisson sampling of reactions that consume reactants can yield negative population numbers, and thus needs to be combined with time step selection and rejection sampling \cite{Cao2005}. 

\subsubsection{Connection to the reaction rate equation}
Tau-leaping is related to the RRE as follows:
If a sufficiently large number of reactions occur within $\Delta t$, i.e. $a_r\left(\vec{X}(t),t\right)\Delta t \gg 1$, then we can approximate the Poisson process by a Gaussian process.
Furthermore, if reactive fluctuations are negligible, i.e. $\sqrt{a_r\left(\vec{X}(t),t\right)\Delta t} \ll a_r\left(\vec{X}(t),t\right)\Delta t$ then we can replace the Gaussian process by its mean value.
It can then be shown \cite{Gillespie2007} that equation~\eqref{eq:tau_leaping} yields

\begin{equation}
  \label{eq:rre}
  \frac{\d\vec{X}}{\dt} = \sum_r\vec{\nu}_ra_{r}\left(\vec{X}(t),t\right),
\end{equation}
which we recognize as the deterministic reaction rate equation for the particle density $\vec{n}(t) = \vec{X}(t)/\Delta V$.
Equation~\eqref{eq:rre} now allows us to identify the usual rate constants from the propensities.
For example, for the bimolecular reaction $S_i + S_j\xrightarrow{k} \varnothing$ the reaction rate constant is $2k/\Delta V$ for $i=j$ and $k/\Delta V$ for $i\neq j$.

\subsubsection{Reaction algorithm outline}
\label{sec:reaction_algorithm}

Complete details regarding the reactive algorithm that we use are found in \cite{Gillespie1977, Cao2005, Cao2006}.
We are interested in advancing $\vec{X}(t)$ from time $t$ to time $t+\Delta t$ for a set of reactions $\vec{R}$.
Letting $\tau$ be the simulated time within $\Delta t$, this proceeds as follows:

\begin{enumerate}
\item Partition $\vec{R}$ into critical and non-critical reaction sets $\vec{R}_{\textrm{c}}$ and $\vec{R}_{\textrm{nc}}$.
  The critical reactions are defined as the set of reactions that are within $N_{\textrm{c}}$ firings of consuming its reactants.
  We take $N_{\textrm{c}} = 5$ in this paper. 
\item Compute all propensities, the total propensity $A$ and the critical propensity $A_{\textrm{c}}$:
  
  \begin{subequations}
    \begin{align}
      A &= \sum_{r\in \vec{R}} a_{r, \tau}, \\
      A_{\textrm{c}} &= \sum_{r\in \vec{R}_c} a_{r, \tau},
    \end{align}
  \end{subequations}
  where $a_{r,\tau} \equiv a_r\left(\vec{X}(t+\tau),t+\tau\right)$.
\item Compute the time $\Delta\tau_{\textrm{c}}$ until the next critical reaction:

  \begin{equation}
    \Delta\tau_{\textrm{c}} = \frac{1}{A_{\textrm{c}}}\ln\left(\frac{1}{u_1}\right),
  \end{equation}
  where $u_1\in[0,1]$ is a uniformly distributed random number. 
\item Compute a permitted time step $\Delta\tau_{\textrm{nc}}$ such that non-critical reaction propensities do not change by a relative factor greater than $\epsilon$:

  \begin{equation}
    \Delta\tau_{\textrm{nc}} = \min_{i\in I_{\textrm{rs}}}\left(\frac{\max\left(\frac{\epsilon X_i}{g_i}, 1\right)}{\left|\xi_i\right|}, \frac{\max\left(\frac{\epsilon X_i}{g_i}, 1\right)^2}{\sigma^2_i}\right),
  \end{equation}
  where $I_{\textrm{rs}}$ is the set of reactant species in $\vec{R}_{\textrm{nc}}$ and
  
  \begin{align}
    \xi_i   &= \sum_{r\in\vec{R}_{\textrm{nc}}}\nu_{ri}a_{r, \tau}, \\
    \sigma_i^2 &= \sum_{r\in\vec{R}_{\textrm{nc}}}\nu_{ri}^2a_{r, \tau}.
  \end{align}
  Here, $\nu_{ri}$ is the state change of $X_i$ due to one firing of reaction $r$.
  The factor $g_i$ depends on the highest order of reaction where the reactant $i$ appears \cite{Cao2005}.
  In this paper we consider only first order reactions and then $g_i=1$. 

\item To halt integration at time $\Delta t$, select a reactive substep $\Delta\tau$ within $\Delta t$ from
  
  \begin{equation}
    \Delta \tau = \min\left[\Delta t - \tau, \min\left(\Delta\tau_{\textrm{c}}, \Delta\tau_{\textrm{nc}}\right)\right].
  \end{equation}

\item Resolve reactions as follows:
  
  \begin{enumerate}
  \item \underline{If $\Delta\tau_{\textrm{c}} < \Delta\tau_{\textrm{nc}}$ and $\Delta \tau_{\textrm{c}} < \Delta t - \tau$:}
    One critical reaction fires.
    Determine the critical reaction $r_c$ from
    
    \begin{equation}
      r_c = \text{smallest integer satisfying} \sum_{r=0}^{r_c-1} a_r> u_2 A_c,      
    \end{equation}
    where $u_2\in[0,1]$ is sampled from a uniform distribution.
    The sum only runs over the critical reactions.
    Advance the state $\vec{X}$ with the results from the SSA and tau-leaping reaction firings:
    
    \begin{align}
      \vec{X} &\rightarrow \vec{X} + \vec{\nu}_{r_c} + \sum_{r\in\vec{R}_{\textrm{nc}}}\mathcal{P}\left(a_{r, \tau}\Delta\tau\right). 
    \end{align}
  \item \underline{Otherwise:}
    No critical reactions fire.
    Advance $\vec{X}$ over $\Delta\tau$ with the non-critical reactions only:
    
    \begin{equation}
      \vec{X} \rightarrow \vec{X} + \sum_{r\in\vec{R}_{\textrm{nc}}}\mathcal{P}\left(a_{r, \tau}\Delta\tau\right).
    \end{equation}
    
    An exception is made if $A\Delta\tau$ is smaller than some factor (we take $A\Delta\tau \leq 1$) since tau-leaping is inefficient in this limit.
    In this case we switch to SSA stepping using the whole reaction network, taking up to $N_{\textrm{SSA}}=10$ steps for a total integration time $\Delta\tau^\prime$.
    Obviously, we restrict this integration to $\Delta\tau^\prime \leq \Delta\tau$.
  \end{enumerate}
\item Check if $\vec{X}$ is a valid state:
  \begin{enumerate}
  \item If any particle numbers in $\vec{X}$ are negative, reject the update. Let $\Delta\tau_{\textrm{nc}} \rightarrow \Delta\tau_{\textrm{nc}}/2$ and return to step 5.
  \item Otherwise, increment $\tau$ by $\Delta\tau$, or by $\Delta\tau^{\prime}$ if triggering use of the SSA in step 6(b).
  \end{enumerate}
\item If $\tau < \Delta t$, return to step 1.
\end{enumerate}

The above algorithm is a well-tested procedure which uses the SSA and tau-leaping algorithms in their respective limits.
The factor $\epsilon$ determines the maximum permitted relative change in the propensities during one tau-leaping step, and therefore adjusts the accuracy and number of reactive substeps that the algorithm will take.

\subsubsection{Comparing reaction algorithms}

To highlight the potential importance of reactive fluctuations, we compare the stochastic reaction algorithm with the RRE for a zero-dimensional test case that illustrates basic electron-neutral interactions in atmospheric pressure air. 
For simplicity we only consider electron impact ionization and attachment, i.e.

\begin{subequations}
  \begin{align}
    \e + \varnothing &\xrightarrow{k_\alpha} \e + \e + \varnothing, \\
    \e + \varnothing &\xrightarrow{k_\eta} \varnothing, 
  \end{align}
\end{subequations}
where the ionization and attachment rates are computed using BOLSIG+ (see section~\ref{sec:examples} for further details).
The breakdown field is $E_b\approx\SI{3}{\kilo\volt\per\milli\meter}$, and the exact solution to the RRE (equation~\eqref{eq:rre}) is

\begin{equation}
  X_{\e}(t) = X_{\e}(0)\exp\left[\left(k_\alpha - k_\eta\right)t\right],
\end{equation}
where $X_{\e}(0)$ is the number of starting electrons.

\begin{figure}[h!t!b!]
  \centering
  \includegraphics{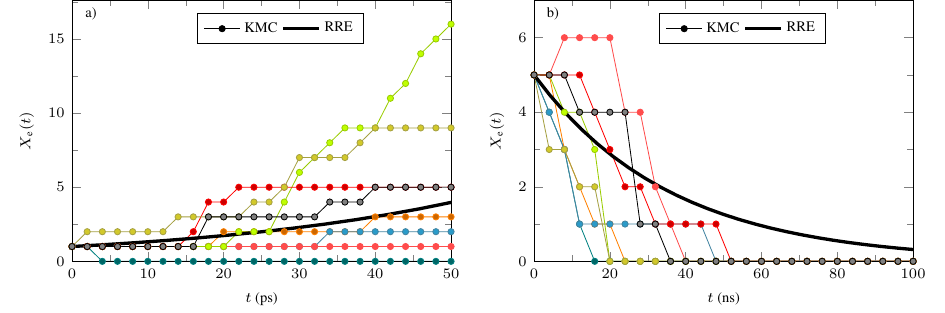}
  \caption{Comparison between the KMC algorithm and the RRE.
    a) Breakdown conditions with $E=\SI{10}{\kilo\volt\per\milli\meter}$ and a single starting electron.
    b) Sub-breakdown conditions with $E = \SI{1.7}{\kilo\volt\per\milli\meter}$ and five initial electrons. }
  \label{fig:AvalancheAttachment}
\end{figure}

Figure~\ref{fig:AvalancheAttachment}a) shows the results under breakdown conditions with an electric field $E=\SI{10}{\kilo\volt\per\milli\meter}$ and a single starting electron.
We have advanced for \SI{50}{\pico\second} which from the RRE yields $4\text{-}5$ electrons.
This value is compared with the predictions of eight independent runs using the stochastic algorithm.
The stochastic algorithm shows considerable variation in the final number of electrons, including one case where the initial electron attached.

Figure~\ref{fig:AvalancheAttachment}b) shows a similar case for sub-breakdown conditions with an electric field $E = \SI{1.7}{\kilo\volt\per\milli\meter}$ and five initial electrons. 
In this case attachment processes dominate the evolution. 
We find that the hybrid algorithm eventually leads to attachment of all five initial electrons while the RRE yields a solution which only asymptotically decays to zero, i.e. it contains fractional electrons.
This latter point is particularly pertinent to positive streamers in highly attaching gases (e.g., sulphur-hexafluoride).
In a fluid model the seed electrons ahead of the streamer never completely attach and form negative ions.
Rather, the asymptotic decay of the electron density means that a computational fraction of the electron is always available for further seeding the streamer, artificially reducing fluctuations at the streamer tip. 

\subsection{Model remarks}

The \ito-KMC model presented above is a microscopic drift-diffusion model with stochastic chemistry, and in the above we have shown that it recovers the standard drift-diffusion-reaction fluid model in the coarse-grained deterministic limit.
In principle, one can think of the \^Ito-KMC model as a non-kinetic PIC method that samples the macroscopic evolution using computational particles that represent average electrons. 
Our model rectifies some shortcomings of macroscopic drift-diffusion models, in particular those that pertain to numerical stability and efficiency, but also by maintaining a particle description in regions where the plasma is rarefied.

The \ito-KMC model is qualitatively similar to the model by \citet{Luque2011}, which is essentially a reaction-diffusion master equation (RDME) model supplemented with electron drift.
The RDME model evolves the total number of particles in a cell using stochastic sampling of transfer rates. 
However, the \ito-KMC model has a few important distinctions. 
Firstly, we expand beyond pure tau-leaping for the plasma chemistry, with the primary benefits being guaranteed non-negativeness and adjustable accuracy.
The same algorithm could be used in the \citet{Luque2011} model. 
Secondly, the RDME model \cite{Luque2011} does not generalize very well to the strong drift regime of streamers where negative transfer probabilities between grid cells can appear \cite{Luque2011, Noel2018}. 
The \ito-KMC model resolves these issues, but the cost is the adoption of a microscopic model rather than a mesoscopic one.

We have not included any energy description for the electrons or ions.
In fact, it is generally not clear if the model can be extended to include energy transport in such a way that one also recovers the fluid electron energy transport equation when coarse-graining the model.
However, an excellent alternative is to combine \ito-KMC with a kinetic electron description. 
The \ito-KMC method is already a particle model and so the inclusion of kinetic electrons is possible, and is most likely algorithmically simpler than existing hybrid models based on fluid-particle couplings \cite{Li2010,Li2012}.

\section{Computer implementation}
\label{sec:implementation}

In this section we present our implementation of the new PIC model on cut-cell Cartesian AMR grids.
The equations of motion are the \ito-KMC-Poisson system

\begin{subequations}
  \label{eq:ito_poisson}
  \begin{align}
    \label{eq:ito_kernel}
    \d\bm{X}_p &= \bm{V}_p\dt + \sqrt{2D_p\dt}\bm{\mathcal{N}}_p,\\
    \label{eq:kmc}
    \vec{X}\left(t\right) &\xrightarrow{\vec{R}} \vec{X}\left(t+\d t\right),\\
    \label{eq:poisson}
    \nabla^2\Phi &= -\frac{\rho}{\epsilon_0},
\end{align}
\end{subequations}
where equation~\eqref{eq:poisson} is the Poisson equation for the electric field $\bm{E} = -\nabla\Phi$ where $\Phi$ is the electrostatic potential and $\rho$ is the space charge density.
This model has been implemented into the chombo-discharge\footnote{\url{https://github.com/chombo-discharge/chombo-discharge}} code, which we have used for streamer simulations in the past \cite{8785920,Marskar2019,Marskar2019b,Marskar2020,Meyer2020,Meyer2022}.

\subsection{Spatial discretization}
\begin{figure}[h!t!b!]
  \centering
  \includegraphics{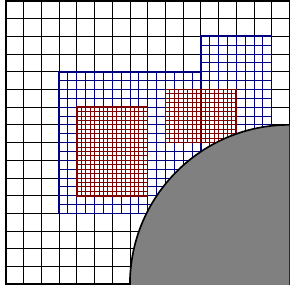}
  \caption{Classic cartoon of a cut-cell AMR grids with a solid boundary (shaded region).
    The coarsest grid covers a domain of $16\times16$ cells with two-levels of properly nested refined grids.
    Bold lines represent grid boundaries.}
  \label{fig:amr}
\end{figure}

We discretize the equations over a Cartesian grid with patch-based AMR, see figure~\ref{fig:amr}.
With AMR, the equations of motion are solved over a hierarchy of grid levels $l\in 0,1, \ldots, l_{\textrm{max}}$. 
When refining a grid level, the resolution increases by a factor two, i.e. $\Delta x_{l+1} = \Delta x_l/2$. 
Each grid level consists of a union of properly nested rectangular grid boxes. 
That is, the valid region of levels $l-1$ and $l+1$ are separated by at least one grid cell on level $l$, and boxes are disjoint (non-overlapping) on each level. 
As in previous publications relating to non-equilibrium gas discharges \cite{Marskar2019, Marskar2019b, 8785920, Marskar2020, Meyer2020, Meyer2022} we use the Chombo \cite{chombo} library for handling the AMR infrastructure. 
Algorithmic details that are not specific to the \ito-KMC model are found in references \cite{Marskar2019, Marskar2019b, 8785920, Marskar2020, chombo}. 

\subsection{Charge deposition \& interpolation}
Deposition of particles and interpolation to the particle positions is done using a cloud-in-cell (CIC) scheme with Cartesian AMR.
The mesh densities $n$ are given on cell centers $n\left(\bm{x}_{\bm{i}}\right) = n_{\bm{i}}$ where $\bm{i}$ is a multi-dimensional index and $\bm{x}_{\bm{i}} = \bm{i}\Delta x$.
The mesh density is given by

\begin{equation}
  n_{\bm{i}}\ = \sum_{p\in\textrm{R}(\bm{i})} \left(\frac{w_p}{\Delta V_{\bm{i}}}\right)\mathcal{W}_{\textrm{cic}}\left(\frac{\bm{x}_{\bm{i}} - \bm{X}_p}{\Delta x}\right),
\end{equation}
where $p\in \textrm{R}(\bm{i})$ indicates particles whose clouds extend into cell $\bm{i}$, $w_p$ is the particle weight, $\Delta V_{\bm{i}}$ is the volume of the grid cell and

\begin{align}
  \mathcal{W}_{\textrm{cic}}\left(\bm{x}\right) &= \prod_{s=1}^dW_{\textrm{cic}}(x_s), \\
  W_{\textrm{cic}}(x) &= \begin{cases}
    1 - |x|, & |x|<1, \\
    0, & \textrm{otherwise}.
  \end{cases}
\end{align}

\begin{figure}[h!t!b!]
  \centering
  \includegraphics{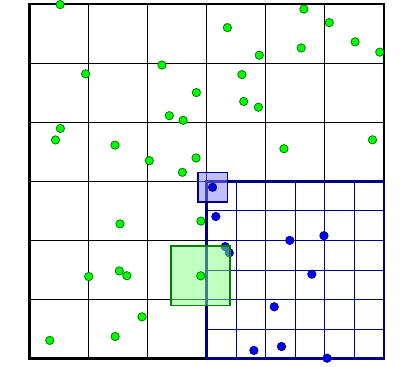}
  \caption{Refinement boundary deposition procedure at the coarse-fine interface between levels $l$ and $l+1$.
    Particles on level $l$ can deposit into level $l+1$; they retain their original particle shape.
    Particles on level $l+1$ can also deposit into level $l$; the deposition weight across the refinement boundary is added to the corresponding coarse-side grid cells.
    }
  \label{fig:deposition}
\end{figure}

Deposition of particles near refinement boundaries are sketched in figure~\ref{fig:deposition} and handled as follows:
If the particle cloud for a particle at level $l+1$ hangs over the refinement boundary into the coarse level $l$, the deposition weight is added to the corresponding level $l$ cells and then normalized by the appropriate volume fraction.
Particles that live on a coarse grid level $l$ but whose clouds hang into level $l+1$ have particle widths $\Delta x_l$ on level $l$, and for factor two refinements they may extend into the first strip of cells on level $l+1$ (see figure~\ref{fig:deposition}).
In order to ensure that this weight ends up in the correct cells on level $l+1$, these particles are also deposited on level $l+1$, but using the original particle width $\Delta x_l$.
These particles thus have a width of $2\Delta x_{l+1}$ and in 3D they can deposit into at most $9$ grid cells on level $l+1$.
Their deposition function on level $l+1$ is

\begin{equation}
  W_{\textrm{cic}}^\prime(x) = \begin{cases}
      \frac{1}{2}, & |x| \leq \frac{1}{2} \\
      \frac{1}{2}\left(\frac{3}{2} - |x|\right), & \frac{1}{2} < |x| \leq \frac{3}{2} \\
      0, & \textrm{otherwise}. 
    \end{cases}
\end{equation}

\subsection{Semi-implicit Euler-Maruyama method}
The standard Euler-Maruyama method for equation~\eqref{eq:ito} is

\begin{equation}
  \bm{X}_p^{k+1} = \bm{X}_p^{k} + \Delta t\bm{V}_p^{k} + \sqrt{2D_p^{k} \Delta t}\bm{\mathcal{N}}_p.
\end{equation}
Coupled to equation~\eqref{eq:poisson}, the discretization must be restricted by the dielectric relaxation time in order to be stable, i.e. $\Delta t \leq \sigma/\epsilon_0$
We remove this limitation by using a semi-implicit formulation as follows:

\begin{equation}
  \label{eq:particle_kernel}
  \bm{X}_p^{k+1} = \bm{X}_p^{k} + \Delta t\bm{V}^{k+1}_p + \sqrt{2D_p^{k} \Delta t}\bm{\mathcal{N}}_ p.
\end{equation}
where

\begin{equation}
  \label{eq:Vk1}
  \bm{V}_p^{k+1} = \sgn\left(Z_s\right)\mu_p^k\bm{E}^{k+1}\left(\bm{X}_p^k\right)
\end{equation}
and $Z_s$ is the charge number for species $s$, and $\sgn$ is the sign operator.
I.e., we have $\sgn\left(Z_s\right) = -1$ for electrons and $\sgn\left(Z_s\right) = 1$ for positive ions. 
We achieve this coupling by first solving the Poisson equation

\begin{equation}
  \nabla\cdot\bm{E}^{k+1} = \frac{\rho^{k+1}}{\epsilon_0},
\end{equation}
which to first order in $\Delta t$ can be written

\begin{equation}
  \label{eq:poisson_implicit}
  \nabla\cdot\bm{E}^{k+1} = \frac{1}{\epsilon_0}\rho^\dagger - \frac{\Delta t}{\epsilon_0}\nabla\cdot\bm{J}_{\textrm{adv.}}^k,
\end{equation}
where $\bm{J}_{\textrm{adv.}}^k$ is the advective current density and $\rho^\dagger$ is the space charge density computed from the update

\begin{equation}
  \label{eq:Xdagger}
  \bm{X}_ p^{\dagger} = \bm{X}_ p^{k} + \sqrt{2D_p^{k} \Delta t}\bm{\mathcal{N}}_p.
\end{equation}
For a species $s$ the advective current density at a grid point $\bm{x}_{\bm{i}}$ is

\begin{equation}
  \label{eq:J_adv}
  \bm{J}_{\bm{i}, \textrm{adv.}}^{s,k} = q_\e\left|Z_s\right|\sum_{p} \mu_s^k\bm{E}^{k+1}\left(\bm{X}_p^k\right)\mathcal{W}\left(\bm{x}_{\bm{i}}-\bm{X}_p^k\right).
\end{equation}
Expanding $\bm{E}^{k+1}\left(\bm{X}_p^{k}\right)$ as a polynomial around the grid point $\bm{x}_{\bm{i}}$ yields

\begin{equation}
  \bm{E}^{k+1}\left(\bm{X}_p^k\right) \approx \bm{E}^{k+1}_{\bm{i}} - \left(\bm{x}_{\bm{i}} - \bm{X}_p^{k}\right)\cdot\nabla\bm{E}^{k+1}_{\bm{i}} + \mathcal{O}\left(\left(\bm{x}_{\bm{i}} - \bm{X}^k_p\right)^2\right),
\end{equation}
where $\bm{E}_{\bm{i}} = \bm{E}(\bm{x}_{\bm{i}})$.
Equation~\eqref{eq:J_adv} yields

\begin{equation}
  \bm{J}_{\bm{i}, \textrm{adv.}}^{s,k} \approx \left[q_\e\left|Z_s\right|\sum_{p}\mu_p^k\mathcal{W}\left(\bm{x}_{\bm{i}}-\bm{X}_p^k\right)\right] \bm{E}^{k+1}_{\bm{i}} -\left[q_\e\left|Z_s\right|\sum_{p} \mu_p^k\mathcal{W}\left(\bm{x}_{\bm{i}}-\bm{X}_p^k\right)\left(\bm{x}_{\bm{i}} - \bm{X}_p^k\right)\right]\cdot\nabla\bm{E}^{k+1}_{\bm{i}},
\end{equation}
where the first term is the conventional Ohmic contribution that we recognize from semi-implicit formulations for fluid models \cite{Ventzek1994}.
The support of $\mathcal{W}$ is $\Delta x$, and so the second term scales as $\Delta x$ as well.
When used together with equation~\eqref{eq:poisson_implicit} this term scales as $\mathcal{O}\left(\Delta x\Delta t\right)$ in the semi-implicit Poisson equation, and it also has a small error constant: 
The moments $\mathcal{W}\left(\bm{x}_{\bm{i}}-\bm{X}_p^k\right)\left(\bm{x}_{\bm{i}}-\bm{X}_p^k\right)$ are anti-symmetric in $\bm{X}_p^k$ with respect to the grid cell center $\bm{x}_{\bm{i}}$, so when particles distribute uniformly over a grid cell the summation yields the zero vector.
This is, for example, the case in the discharge channels where there are many electrons per grid cell, whereas outside of the channel the current is negligibly small.  
Summing over all species $s$ to leading order yields

\begin{equation}
  \label{eq:J_advective}
  \bm{J}^{k}_{\textrm{adv.}} = \left(q_\e\sum_s\left|Z_s\right|\mu_s^kn_s^k\right)\bm{E}^{k+1},
\end{equation}
where $\mu_s^k$ and $n_s^k$ are mesh variables for species $s$ (we have suppressed the index $\bm{i}$). 
Thus, ignoring higher-order moments, equation~\eqref{eq:poisson_implicit} can be written in the familiar form \cite{Ventzek1994}

\begin{equation}
  \label{eq:poisson_simp}
  \nabla\cdot\left[\left(1 + \frac{\sigma^k\Delta t}{\epsilon_0}\right)\bm{E}^{k+1}\right] = \frac{1}{\epsilon_0}\rho^\dagger,
\end{equation}
where

\begin{equation}
  \label{eq:conductivity}
  \sigma^k = q_\e\sum_s\left|Z_s\right|\mu_s^{k}n_s^k
\end{equation}
is the conductivity of the plasma.

Equation~\eqref{eq:poisson_simp} is discretized with finite volumes, using a standard 9-point stencil in the interior points (and flux matching at the coarse-fine interface).
The embedded boundary fluxes are also constructed to second order, using additional interior points when evaluating the normal derivative on the cut-cell boundary centroids. 
The corresponding linear system is solved using geometric multigrid with V-cycling, using red-black Gauss-Seidel relaxation as the smoother on each grid level and a biconjugate gradient stabilized method (BiCGSTAB) as a bottom solver.
A relative exit tolerance of $10^{-10}$ is used as a convergence criterion for multigrid. 
Further details regarding the finite volume discretization of the variable-coefficient Poisson equation and its embedding into geometric multigrid in the presence of embedded boundaries and Cartesian AMR are given in e.g. \cite{chombo, Johansen1998, McCorquodale2001, Schwartz2006}.
After obtaining the electric field $\bm{E}^{k+1}$ we compute $\bm{V}_p^{k+1}$ from equation~\eqref{eq:Vk1} and complete the particle update (equation~\eqref{eq:particle_kernel}).

\subsection{KMC-particle coupling}
\label{sec:kmc_particle}

The reaction algorithm  solves for the total number of particles in a grid cell and leaves substantial freedom in how one assigns the chemistry products into new computational particles.
Since we are concerned with methods that potentially use very large time steps, the creation of computational particles with physical weights $w=1$ is not possible due to the large number of physical particles generated in a time step.
In this paper, if the reaction step led to net creation of particles we instead create at most $N_{\ppc}^{\text{new}}=64$ new computational particles in the cell. 
If the KMC solver gave $N > N_{\ppc}^{\text{new}}$ physical particles in the cell, we construct $N_{\ppc}^{\textrm{new}}$ computational particles with weights

\begin{equation}
  \label{eq:w}
  w = N\div N_{\ppc}^{\text{new}},
\end{equation}
where $\div$ denotes integer division.
The remainder $\mod(N, N_{\ppc}^{\text{new}})$ is assigned to one of the new particles.
These particles are later merged with the computational particles that already exist in the cell.
If the KMC algorithm led to net loss of particles, we remove the weight directly from the existing computational particles.

Production of particles in cut-cells only takes place in the valid region of the cell. 
Particle positions $\bm{X}_p$ are drawn from a uniform distribution in each coordinate direction with the requirement

\begin{equation}
  \left(\bm{X}_p-\bm{x}_c\right)\cdot \bm{\hat{n}}_c \geq 0,
\end{equation}
where $\bm{x}_c$ is the cut-cell boundary centroid and $\bm{\hat{n}}_c$ is the cut-cell boundary normal.
Since cut-cell volume fractions can be arbitrarily small, we optimize this step by only drawing the position $\bm{X}_p$ inside the minimum bounding box that encloses the valid region of the cut-cell.

\subsection{Photon generation and transport}
Photons are also treated with a particle method, and for simplicity we consider instantaneous transport where we don't have to track the photons in time.
As with the particles, the KMC algorithm provides the number of physical photons that is generated in the reaction step, which we limit to $N_{\ppc}^{\text{new}}=64$ computational photons.
Photon weights and emission positions are assigned in the same way as we do for the particles.

Photon absorption positions are determined individually for each (super-)photon.
For example, assume that a photon has some frequency $f$ and is emitted from an initial position $\bm{Y}_{f}^0$.
This photon is absorbed at position

\begin{equation}
  \label{eq:photon_transport}
  \bm{Y}_f = \bm{Y}_{f}^0 + r_f \bm{\hat{c}},
\end{equation}
where $r_f$ is a random number drawn from an exponential distribution with parameter $\kappa(f)$, and $\bm{\hat{c}}$ is a uniformly distributed random point on the unit sphere.
Here, $\kappa(f)$ is the mean absorption coefficient in the gas for a photon with frequency $f$, i.e. $1/\kappa(f)$ is the mean absorption length.
Since specral absorption lines also have a spectral width, the mean absorption coefficient $\kappa(f)$ is frequency dependent.
When we sample the photon generation, we begin by sampling the spectral line by stochastically determining $f$ according to some distribution.
We then use a known expression for $\kappa(f)$ for stochastically determining $r_f$ for each photon.
Depending on the photoionization model that is used, one can sample multiple spectral lines \cite{Stephens2018} in combination \cite{Chanrion2008} or individually \cite{Marskar2020}.
Photoemission is disregarded in this paper, so if a photon trajectory intersects an internal boundary (e.g. an electrode) or a domain boundary, it is removed from the simulation.

\subsection{Superparticle management}
\label{sec:superparticles}

In order to maintain a manageable number of particles, only computational particles that represent many physical particles are tracked.
Our particle merging and splitting strategy uses a bounding volume hierarchy with $k\text{-}d$ trees for locating spatial clusters of particles, and particles are merged/split within each cluster.
We use a standard tree structure which uses top-down construction, i.e. it is hierarchically built from the root node and downwards.
However, the algorithm that we use for splitting a leaf node is new and it is therefore discussed in detail.

\begin{figure}[h!t!b!]
  \centering
  \includegraphics{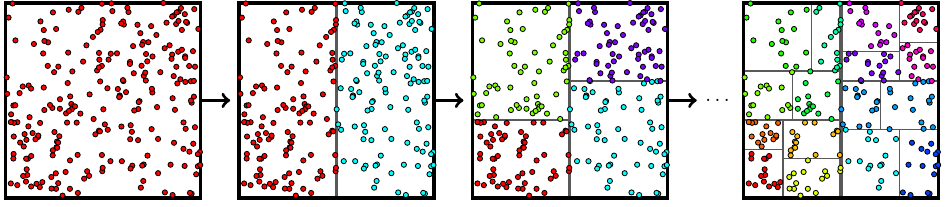}
  \caption{Concept sketch of Cartesian 2D bounding volume hierarchy generation.
    Solid lines inside the cell indicate splitting lines (planes in 3D).
    From left to right: Original particles, two bounding volumes, four bounding volumes, and 16 bounding volumes. }
  \label{fig:LeafSplitting}
\end{figure}

Initially, a leaf node $L$ contains a list $\vec{P}_L$ of $M$ particles that are each identified by a tuple $\langle\bm{X}_p, w_p\rangle$, where $w_p$ is the particle weight. 
Then, $\vec{P}_L$ is sorted based on one of the axis coordinates and split into two bounding volumes such that the total weight of the two halves differ by at most one physical particle. 
This process is shown in figure~\ref{fig:LeafSplitting} and proceeds as follows:
\begin{enumerate}
\item Pick a splitting direction.
  We choose the coordinate direction where the minimum bounding box enclosing the particles has the largest extent. 
\item Sort the particle list $\vec{P}_L$ from smallest to largest coordinate in splitting direction. 
\item Locate the median particle with index $p^\prime$ in the list, where $p^\prime > 1$ is the smallest index satisfying
  
  \begin{equation}
    \label{eq:weight_median}
    \sum_{p=1}^{p^\prime-1} w_p + w_{p^\prime} > \sum_{p=p^\prime+1}^M w_p.
  \end{equation}
  The index $p^\prime$ indicates the position of the particle on the splitting plane, i.e. all particles $p < p^\prime$ are found on the left hand side of the splitting plane and all particles $p > p^\prime$ are found on the right-hand side .
\item Transfer particles $p\in[1, p^\prime-1]$ to a new list $\vec{P}_l$ in the left leaf node, and particles $p\in[p^\prime+1,M]$ to another list $\vec{P}_r$ in the right leaf node.
\item Assign the median particle $p^\prime$:
  \begin{enumerate}
  \item If the median particle is a physical particle the particle is assigned to whichever child list ($\vec{P}_l$ or $\vec{P}_r$) has the lowest total weight. 
  \item If the median particle is a superparticle, i.e. $w_{p^\prime} \geq 2$, it is split into two new particles with the same position $\bm{X}_{p^\prime}$ but with new weights.
    Due to the median selection in equation~\eqref{eq:weight_median}, these weights can be constructed such that the weight of $\vec{P}_l$ and $\vec{P}_r$ differ by at most one physical particle.     
  \end{enumerate}

\end{enumerate}

Because we merge particles by groups rather than in pairs \cite{Teunissen2014}, the process above only proceeds until we have $N_{\ppc}$ leaves in the tree.
Choosing the final number of computational particles to be a factor of two gives a balanced tree where all the leaves exist on the same tree level, and in this case the number of physical particles between any two arbitrary leaves in the tree differs by at most one.
At the end of the tree-building algorithm each leaf node represents a bounding volume with a list $\vec{P}$ of computational particles.
The particles in this list become a new superparticle with weight and position

\begin{subequations}
  \begin{align}
    w &= \sum_{p\in\vec{P}} w_p\\
    \bm{X} &= \frac{1}{w}\sum_{p\in\vec{P}} w_p\bm{X}_p. 
  \end{align}
\end{subequations}
Particle merging is done on a cell-by-cell basis in order to prevent creation of particles that lie inside the embedded boundary.
The algorithm also handles splitting of superparticles.
If a cell contains a single particle with a large weight then step (v) in the above algorithm ensures that this particle is hierarchically split until we have created $N_{\ppc}$ new particles.

\begin{figure}[h!t!b!]
  \centering
  \includegraphics{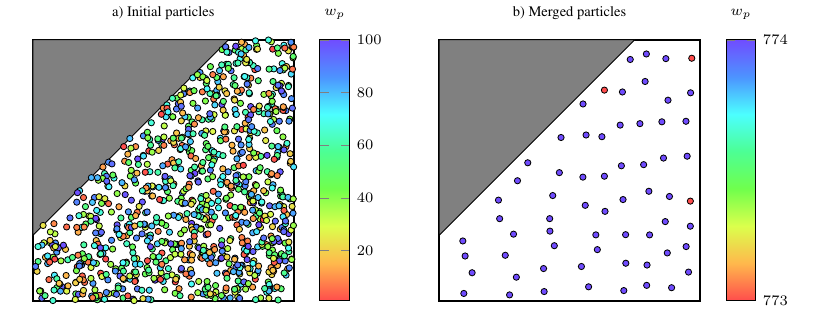}
  \caption{
    Particle merging example.
    a) Initial particles.
    b) Merged particles with $N_{\ppc}=64$.
    The labels on the colorbar indicate particle weights $w_p$.
  }
  \label{fig:BVH}
\end{figure}

Figure~\ref{fig:BVH} shows an example of merging $1000$ initial particles whose positions are uniformly distributed within a cut-cell and whose weights are uniformly distributed on the interval $w_p\in[1,100]$.
These particles are merged into $N_{\ppc} = 64$ particles, and as expected the final particle weights differ by at most one.

\subsection{Final algorithm}
The final algorithm for integration over a time step $\Delta t$ is as follows:
\begin{enumerate}
\item Compute the conductivity $\sigma^k$.
\item Perform the diffusive advance:
  
  \begin{displaymath}
    \bm{X}_p^\dagger = \bm{X}_p^k + \sqrt{2D_p^k\Delta t}\bm{\mathcal{N}}_p.
  \end{displaymath}
\item Compute the space charge density $\rho^\dagger = \rho\left(\bm{X}_p^\dagger\right)$ and solve for $\bm{E}^{k+1}$ using equation~\eqref{eq:poisson_simp}.
\item Interpolate particle velocities $\bm{V}_p^{k+1} = \bm{v}\left[\bm{E}^{k+1}\left(\bm{X}_p^{k}\right)\right]$.
\item Advect particles $\bm{X}_p^{k+1} = \bm{X}_p^\dagger + \bm{V}_p^{k+1}\Delta t$. 
\item Move photons $\bm{Y} = \bm{Y}_0 + r_\kappa \bm{\hat{c}}$ using equation~\eqref{eq:photon_transport}.
\item Advance the reaction network over $\Delta t$, see section~\ref{sec:plasma_chemistry} and section~\ref{sec:kmc_particle}.
\item Manage superparticles, section~\ref{sec:superparticles}.
\end{enumerate}

Conceptually, the above algorithm uses a Godunov splitting between particle transport (steps 1 through 5) and plasma chemistry (step 7).
The particle transport step is a first-order accurate semi-implicit discretization, and the plasma chemistry is solved with a stochastic reaction algorithm with adjustable accuracy through the factor $\epsilon$ (section~\ref{sec:plasma_chemistry}).
Setting $\epsilon=\infty$ will accept any tau-leaping step, i.e. the chemistry is resolved with a time step $\Delta t$.
On the other hand, setting $0 < \epsilon \ll 1$ yields a highly accurate chemistry algorithm which may potentially take many substeps within $\Delta t$.
But if high order chemistry is used together with a large splitting step $\Delta t$, the overall stability of the algorithm can deteriorate.
The reason for this is that when transport and field updates are performed rarely but the chemistry integration is highly accurate, the number of free electrons that are generated in a grid cell is overestimated. 
This issue is not unique for \ito-KMC but also occurs for deterministic fluid models when using operator splitting methods with large splitting steps.
Since we use fixed time steps in this paper, we therefore resolve the transport and chemistry with the same time step, i.e. we use $\epsilon = \infty$, and rely on the SSA steps primarily to avoid negative particle numbers.
In the future, we will be extending our methodology to dynamic time stepping (either CFL or physics based), at which point we will be able to leverage the adjustable accuracy features in the KMC integrator.

\subsection{Parallelization}
Our computer implementation is parallelized with flat MPI, using the natural domain decomposition offered by the AMR grids where each MPI rank solves for a subset of the grids on each level.
The simulations are performed away from the strong scaling limit, which left room for load balancing our application.

\begin{figure}[h!t!b!]
  \centering
  \includegraphics{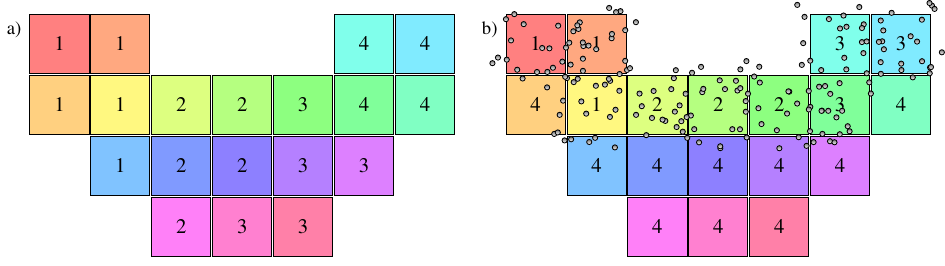}
  \caption{Example of dual mesh load balancing for different kernels. Each colored square represents a grid patch (of e.g. $16^3$ cells).
    The MPI rank ownership is indicated by numbers inscribed in each square.
    a) Example MPI rank assignment and load distribution for kernels whose load scale with the number of grid points.
    b) Example MPI rank assignment and load distribution for kernels that scale with the number of particles. }
  \label{fig:DualMesh}
\end{figure}

The field and particle updates have different computational metrics.
A reasonable proxy for the computational load of the discretized Poisson equation is the number of grid cells in a grid patch, while for the particles the load is better estimated by the number of particles that are assigned to the patch.
We have load balanced our simulations with dual grids.
In this approach we use two sets of AMR grids where the grid levels consist of the same grid patches, but where the assignment of grid subsets among the MPI ranks differ, see figure~\ref{fig:DualMesh}.
One AMR grid set is load balanced with the grid patch volume as a proxy for the computational load, and is used for grid kernels that scale with the number of grid points, e.g. the discretized Poisson equation or advancing the reaction network.
On the other grid, we advance kernels that scale with the number of grid particles, i.e. transport kernels, mesh deposition and interpolation, and superparticle handling.
The dual grid approach adds some computational complexity, but these drawbacks were offset by reductions in simulation times which were up to \SI{40}{\percent}.

\section{Numerical tests}
\label{sec:examples}

\subsection{Simulation conditions}

\begin{figure}[h!t!b!]
  \centering
  \includegraphics{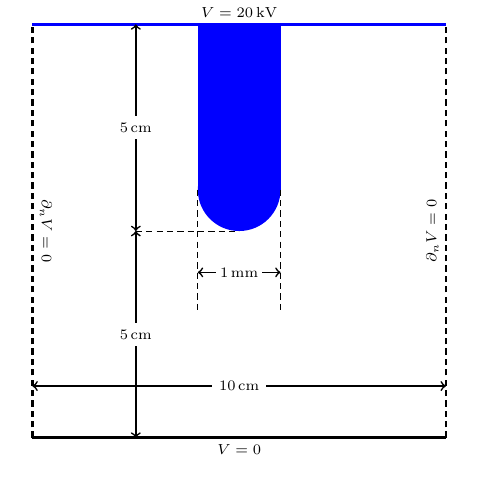}
  \caption{Cross-section of the simulation domain, also showing the electrostatic boundary conditions.
    The dimensions are not to scale.
  }
  \label{fig:domain}
\end{figure}

We consider a \SI{10}{\cubic\centi\meter} computational domain with a vertical needle-plane gap.
A cross section of the computational domain and the boundary conditions is shown schematically in figure~\ref{fig:domain}.
A \SI{5}{\centi\meter} long cylindrical electrode with a spherical cap at the end sticks out of the live electrode plane, and the opposite plane is grounded.
The electrode diameter is \SI{1}{mm} and the vertical distance between the live electrode and the ground plane is \SI{5}{\centi\meter}.
Homogeneous Neumann boundary conditions are used for the Poisson equation on the side faces, and all simulations start from a step voltage of \SI{20}{\kilo\volt}.
The peak initial electric field magnitude is roughly \SI{11}{\kilo\volt\per\milli\meter} on the anode tip.
All simulations use a coarsest AMR level of $\num{128}^3$ cells, but use up to another eight levels of refinement, i.e. up to effective domains of $\num{32768}^3$ cells.

We use a three-species model for discharges in air, consisting of electrons, positive ions, and negative ions.
The plasma kinetics that we use is summarized in table~\ref{tab:air_reactions}.
We focus on using a simple and well-known reaction set for our example simulations.
Using more elaborate plasma chemistry is possible, but not required for our simulation examples. 
The electron diffusion coefficient $D_\e$, mobility $\mu_\e$, temperature $T_{\e}$, ionization frequency $k_\alpha$, and attachment frequency $k_\eta$ are field-dependent and are computed by using BOLSIG+ \cite{Hagelaar2005a} and the SIGLO database \cite{SigloDB}.
The electron-ion and ion-ion recombination rates are

\begin{align}
  k_{\textrm{ep}} &= \beta_{\textrm{ep}}/\Delta V, \\
  k_{\textrm{np}} &= \beta_{\textrm{np}}/\Delta V,
\end{align}
where $\Delta V$ is the grid cell volume and

\begin{align}
  \beta_{\textrm{ep}} &= 1.138\times10^{-11}T_\e^{-0.7}\,\si{\cubic\meter\per\second}, \\
  \beta_{\textrm{np}} &= 2\times10^{-13}\left(300/T\right)^{0.5}\,\si{\cubic\meter\per\second},
\end{align}
where $T=\SI{300}{\kelvin}$ is the gas temperature and $T_\e = T_\e(E)$ is the electron temperature.


\begin{table}[h!t!b!]
  \centering
  \caption{
    Simplified air plasma chemistry used for the example simulations.
    The notation $\varnothing$ indicates an untracked species (e.g., \ce{N_2} or \ce{O_2}) incorporated directly into the rate constant for the reaction.
  }
  \label{tab:air_reactions}  
  \begin{tabular}{lllll}
    \hline
    Reaction & Rate & Propensity & Ref. \\
    \hline
    $\e + \varnothing \rightarrow \e + \e + \M^+$  & $k_\alpha(E)$ & $k_\alpha X_{\e}$ & \cite{Hagelaar2005a} \\
    $\e + \varnothing \rightarrow \M^-$  & $k_\eta(E)$ & $k_\eta X_{\e}(E)$ & \cite{Hagelaar2005a} \\
    $\e + \M^+ \rightarrow \varnothing$  & $k_{\textrm{ep}}(E)$ & $k_{\textrm{ep}}X_{\e}X_{\M^+}$ & \cite{Zhao1995} \\
    $\M^- + \M^+ \rightarrow \varnothing$  & $k_{\textrm{np}}$ & $k_{\text{np}}X_{\M^+}X_{\M^-}$ & \cite{Kossyi1992} \\
    $\e + \varnothing \rightarrow \e + \gamma + \varnothing$ & $k_\gamma(E)$ & $k_\gamma X_{\e}$ & \cite{1982TepVT..20..423Z,Pancheshnyi2015}\\
    \hline      
  \end{tabular}
\end{table}

We use the Zheleznyak photoionization model \cite{1982TepVT..20..423Z} including the corrections by \citet{Pancheshnyi2015} for modeling photon transport for the reaction $\e + \varnothing \xrightarrow{k_\gamma} \e + \gamma + \varnothing$.
The rate constant is

\begin{equation}
  k_\gamma = \frac{p_q}{p + p_q}\nu_Z(E)k_\alpha,
\end{equation}
where $\nu_Z(E)$ is a lumped function that accounts for excitation efficiencies and photoionization probabilities \cite{Pancheshnyi2015}. 
The quenching pressure is $p_q = \SI{40}{\milli\bar}$ and the gas pressure is $p=\SI{1}{\bar}$. 
When a photon is generated within the reaction step we draw a random absorption coefficient as

\begin{equation}
  \kappa_f = K_1\left(\frac{K_2}{K_1}\right)^{\frac{f-f_1}{f_2-f_1}},
\end{equation}
where $K_1 = \SI[per-mode=power]{530}{\per\meter}$, $K_2=\SI[per-mode=power]{3E4}{\per\meter}$, $f_1 = \SI{2.925}{\peta\hertz}$, $f_2 = \SI{3.059}{\peta\hertz}$, and $f$ is a random number sampled from a uniform distribution on the interval $[f_1, f_2]$.
The propagation distance of each photon is then determined by drawing a random number from an exponential distribution with parameter $\kappa_f$.

All simulations start by drawing $10^4$ initial electron-ion pairs uniformly distributed in a sphere with a \SI{500}{\micro\meter} radius centered at the needle tip.
Electron-ion pairs whose positions end up inside the electrode are removed before the simulation starts.

In the computer simulations we refine cells on level $l$ if

\begin{equation}
  \alpha\Delta x_l \geq 1,
\end{equation}
and coarsen if

\begin{equation}
  \alpha\Delta x_l \leq 0.2,
\end{equation} 
where $\Delta x_l$ is the grid spacing on level $l$ and $\alpha$ is the Townsend ionization coefficient.

\subsection{Comparison with hydrodynamics}
In this section we present a comparison  between the \ito-KMC model and an equivalent drift-diffusion-reaction model based on deterministic hydrodynamics (equation~\eqref{eq:sadr} without the stochastic term), using the same photoionization model and transport data.
Since we use cut-cell Cartesian AMR grids, the discretization of the fluid model is bit involved, and is therefore not discussed in detail here.
We follow the discretization that we used in \cite{Marskar2020}.
There, we used a Godunov splitting between plasma transport and reactions, and employed a CTU scheme \cite{Colella1990} that also include transverse slopes in the advective term, permitting a softer CFL constraint.
Here, the only major difference between that discretization and the current one is that we here use explicit diffusion and a semi-implicit coupling to the electric field.

To initialize the fluid model we include the same particle distribution as for the \ito-KMC model and deposit the particles as a density using a nearest-grid-point scheme when the simulation begins.
In the Cartesian 2D comparison we have also raised the potential on the electrode to \SI{80}{\kilo\volt} in order to facilitate propagation of the streamer.
The Cartesian 2D version is included because the lack of fluctuations leads to a single streamer, and the models can then be both qualitatively and quantitatively compared.

\begin{figure}[h!t!b!]
  \centering
  \includegraphics{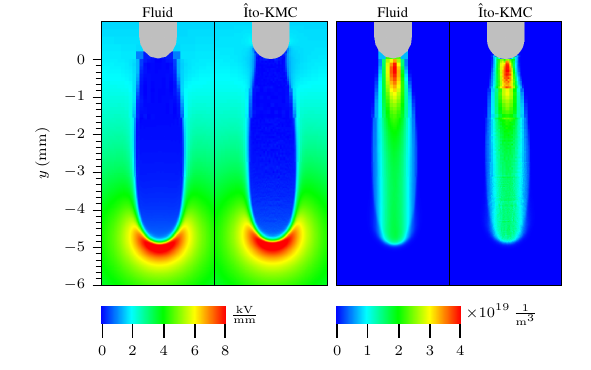}
  \caption{
    Comparison of the field distribution and electron density in the hydrodynamic and \ito-KMC descriptions.
    Left: Field magnitude.
    Right: Electron density.
    }
  \label{fig:HydroComparison}
\end{figure}

\begin{figure}[h!t!b!]
  \centering
  \includegraphics{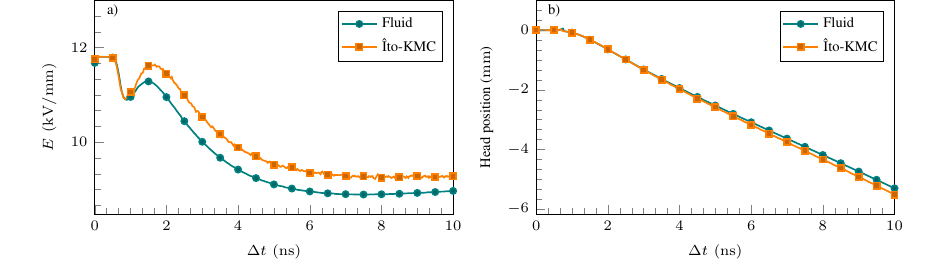}
  \caption{Comparison of the streamer head electric field and streamer head position in the hydrodynamic and \ito-KMC descriptions.}
  \label{fig:HydroItoComparison2D}
\end{figure}

We compare the two models in 2D planar coordinates, using fixed time steps of $\Delta t = \SI{5}{\pico\second}$ for a total integration time of $t = \SI{10}{\nano\second}$.
The field magnitude and electron density after \SI{10}{\nano\second} are shown in figure~\ref{fig:HydroComparison}.
We find that the two solutions are qualitatively very similar (they both tilt slightly to the right due to the initial particle distribution).
Figure~\ref{fig:HydroItoComparison2D} shows the temporal evolution of the maximal electric field and the streamer head position for the simulations.
The head position is defined as the position where the electric field is at its maximum.
Only minor differences are found between the two models:
For example, the largest difference between the maximum electric field in the two models is about \SI{4}{\percent}, while the average streamer velocities agree to within \SI{0.01}{\milli\meter\per\nano\second}.

\subsection{Discrete particle noise}
\label{sec:DiscreteNoise}
\begin{figure}[h!t!b!]
  \centering
  \includegraphics{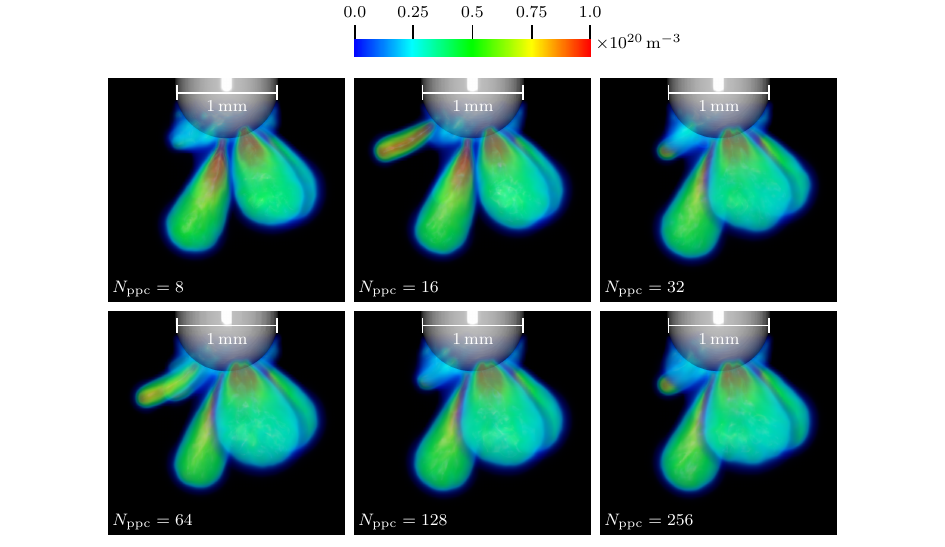}
  \caption{
    Electron density in the neighborhood of the rod electrode computed using various thresholds for the maximum number of computational particles, indicated in each frame by $N_{\ppc}$.
  }
  \label{fig:ParticleNoise}
\end{figure}

To determine if particle noise impacts the simulations, we consider three-dimensional numerical solutions obtained using a varying number of $N_\ppc$ but fixed $\Delta x$ and $\Delta t$.
We select $\Delta x \approx \SI{12}{\micro\meter}$ and $\Delta t = \SI{20}{\pico\second}$ and integrate for \SI{5}{\nano\second}.
The same initial particles are used in these tests.
Figure~\ref{fig:ParticleNoise} shows the resulting electron density in the neighborhood of the electrode using between $8$ and $256$ computational particles per cell (per species).
There is no apparent disagreement or numerical artifacts for these simulations, which suggests that even $N_\ppc=16$ is sufficiently accurate for these particular simulations.
Similar results were obtained by \citet{Teunissen2016}.
This finding can not be extrapolated to coarser grids because if we keep $N_\ppc$ fixed and reduce $\Delta x$ by a factor of two, the average particle weights increase by a factor of \num{8}.
It also warrants mention that there is no particle noise in the KMC algorithm itself, as it operates with the number of physical particles.
However, elevated particle noise still arises due to transport and splitting/merging of superparticles.

\subsection{Grid sensitivity}
\label{sec:convergence}
In this section we perform a grid sensitivity study by varying the spatial and temporal resolutions.
We set $N_\ppc=32$ and consider temporal resolutions ranging from \SIrange{5}{80}{\pico\second}, and spatial resolutions ranging from \SIrange{3}{195}{\micro\meter}, and integrate for \SI{50}{\nano\second}.
Figure~\ref{fig:ItoConvergence} shows the final state for the 35 different simulations, and we make several observations:

\begin{figure}[h!t!b!]
  \centering
  \includegraphics{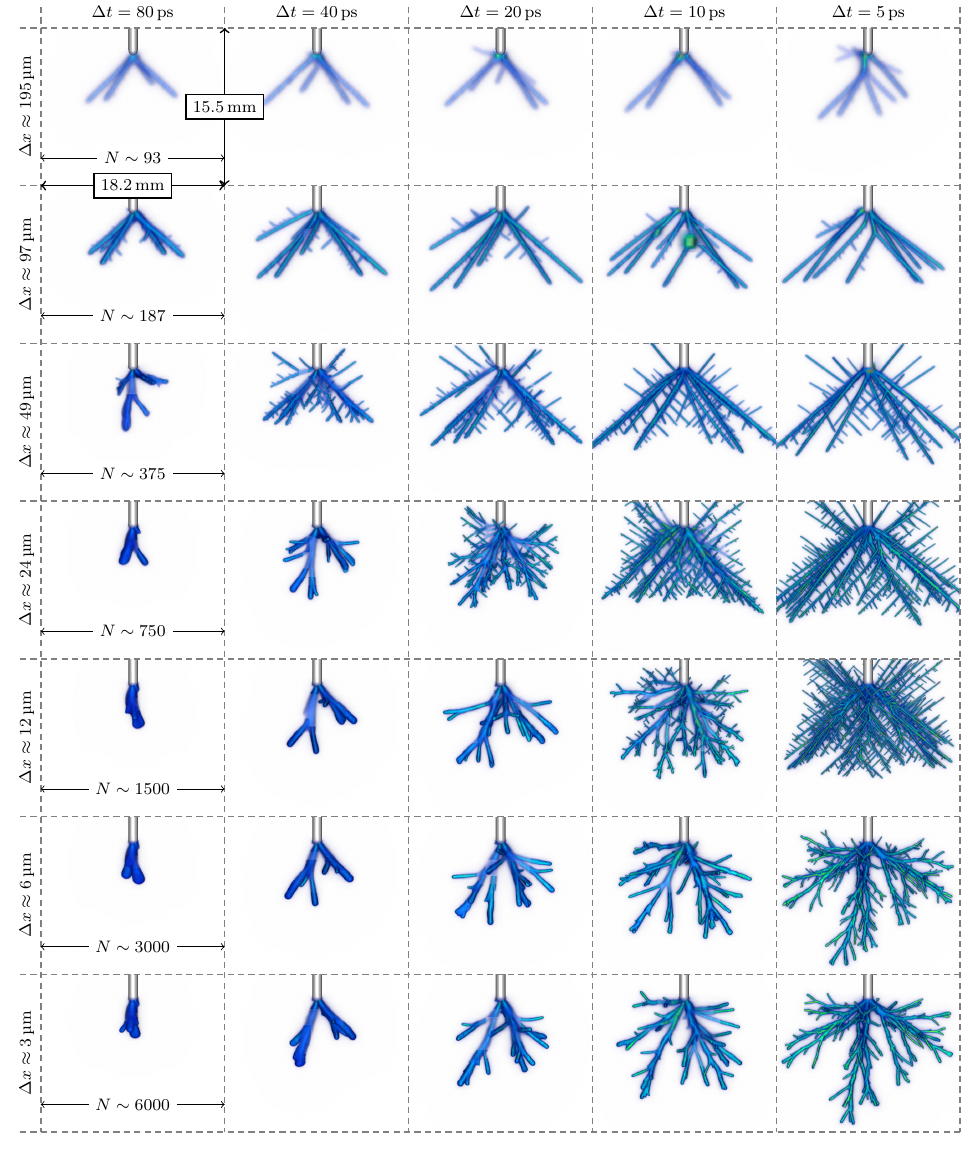}
  \caption{
    Simulation outputs after integrating for $\SI{50}{\nano\second}$.
    The rod diameter is \SI{1}{\milli\meter} and the spatial scale is otherwise indicate in the top left and bottom right frames.
    Along the first column we include the effective number of grid cells along the indicated spatial direction (indicated by $N$).
  }
  \label{fig:ItoConvergence}
\end{figure}

\begin{enumerate}
\item Simulations are stable for all time steps, also when the time step is orders of magnitude larger than the dielectric relaxation time. 
  For the simulations with $\Delta t=\SI{80}{\pico\second}$ and $\Delta x \approx \SI{3}{\micro\meter}$ the time step is equivalent to using an advective CFL number of $> 30$.
  
\item Bounded solutions are obtained for all spatial resolutions, i.e. we do not have $n_\e\rightarrow \infty$ or $E\rightarrow \infty$ for any $\Delta x$ we investigate.
  Here, the spatial resolutions range from moderately fine to extremely coarse.
  From experience with fluid simulations \cite{Marskar2019, Marskar2019b, Marskar2020} we have found that instabilities can occur for branching streamers, particularly if some of the branches become stagnant.

\item Numerical branching occurs on too coarse grids, which can be seen on the column $\Delta t = \SI{80}{\pico\second}$ where the streamer initially splits into four branches, but these four initial branches disappear on finer grids.
  Similar phemonena are seen on the row $\Delta x\approx\SI{24}{\micro\meter}$ where we find small protrusion needles for $\Delta t \leq \SI{20}{\pico\second}$, and similarly for $\Delta x \approx \SI{12}{\micro\meter}$ for $\Delta t \leq \SI{10}{\pico\second}$.
  We believe that these branches appear due to spatial underresolution since for $\Delta x\approx \SI{12}{\micro\meter}$ we only have about 5-10 grid cells for resolving the cross section of the streamers with the smallest radii.

\item The degree of branching decreases when larger time steps are used, which is particularly evident for the row $\Delta x\approx \SI{3}{\micro\meter}$.
  This can be understood in terms of the photoionization-induced noise ahead of the streamer.
  \citet{Bagheri_2019} and \citet{Marskar2020} have shown that the branching behavior depends on the amount of photoionization ahead of the streamer.
  Increasing the amount of photoionization reduces noise in the plasma density ahead of the streamer, which also reduces the amount of branching.
  As larger time steps are used there are more photoelectrons generated during time steps, which artificially suppresses fine-grained temporal variations in the plasma density ahead of the streamer.
  
\item The velocity of the streamers increase with increasing temporal resolution.
  There are at least two reasons for this:

  \begin{enumerate}
  \item The number of electrons in the ionization zone in the streamer grows exponentially, and larger time steps lead to numerical underestimation of the electron impact ionization in the streamer tip.    
  \item When large time steps are used the electrons in the reaction zones at the tip of the streamers can be moved completely out of it.
    For example, the simulation with $\Delta x\approx \SI{3}{\micro\meter}$ and $\Delta t=\SI{80}{\pico\second}$ used an effective CFL number of $>30$.
    The reaction zone is just a few grid cells thick, and moving the electrons too far out of the reaction zone reduces the amount of ionization in it, and thus also velocity of the streamer.
  \end{enumerate}

\item The velocities of the streamers increase slightly when the resolution increases, which we can see on the column $\Delta t = \SI{10}{\pico\second}$.
  We believe this occurs because coarse grids lead to under-resolution of the electric field ahead of the streamer.
  For coarser $\Delta x$ the electric field on the tips is therefore lower, and this reduces the amount of ionization.

\item Streamer radii agree with experimental observations only on the fine grids ($\Delta x \lesssim \SI{6}{\micro\meter}$).
  \citet{Briels2008} have measured streamer diameters in atmospheric air and found that diameters range from \SI{100}{\micro\meter} to \SI{3}{\milli\meter}, depending on experimental conditions like applied voltage, gap inhomogeneity, and various other factors. 
  However, for $\Delta x=\SIrange{97}{195}{\micro\meter}$ we only find streamers with radii $R\sim\SI{1}{\milli\meter}$.
  On finer grids streamers with smaller radii also emerge.
  E.g. for $\Delta x\approx \SI{3}{\micro\meter}$ we find streamer radii ranging from \SI{100}{\micro\meter} to \SI{1}{\milli\meter}, which agree with experimental observations.

\item The electric field at the streamer tips vary by streamer radius.
  On the finest grid $\Delta x\sim \SI{3}{\micro\meter}$ we find $E\sim \SIrange{25}{30}{\kilo\volt\per\milli\meter}$ for filaments with radii on the order of  $R\sim \SIrange{50}{150}{\micro\meter}$. 
  This is in agreement with fluid simulations of positive streamers \cite{Marskar2020} as well as analytical estimates \cite{Chen2013}.
  On grids $\Delta x \geq \SI{24}{\micro\meter}$ we find that the electric field strength at the streamer tips is $E\lesssim \SI{10}{\kilo\volt\per\milli\meter}$, but these solutions are quite clearly underresolved and thus have no practical relevance. 
\end{enumerate}

In summary, we find that the grid resolution should be $\Delta x \leq \SI{6}{\micro\meter}$, which is about the same requirement as in fluid models \cite{Marskar2020}.
The time step should be $\Delta t\leq\SI{10}{\pico\second}$, which at $\Delta x \approx \SI{6}{\micro\meter}$ is about a factor of 5 larger than that permitted through a conventional CFL condition like equation~\eqref{eq:explicitDt}, as shown in figure~\ref{fig:FluidTimeStep}.
We also point out that the \ito-KMC method can maintain this time step even for finer grids, which is not possible for explicit fluid codes.
Although \ito-KMC is quite forgiving for larger time steps, it is clear from figure~\ref{fig:ItoConvergence} that lack of temporal resolution leads to suppression of several morphological features in the discharge, while underresolved grids lead to numerical branching.
Also note that the the fastest time scales in the \ito-KMC method are the same as in fluid methods, which for the reaction set in table~\ref{tab:air_reactions} is the electron ionization impact frequency $\alpha \mu_\e E$.
We have not been able to run numerical convergence tests due to the stochastic component that is involved, which would require ensemble studies at high spatial and temporal resolutions.
We nonetheless observe that the solutions convergence to physically meaningful solutions with streamer diameters and velocities that quantitatively agree with experimental observations \cite{Briels2008}.
Section~\ref{sec:DiscreteNoise} showed that that discrete particle noise had a comparatively low qualitative impact on the simulations, so the artificial branching for $\Delta x\gtrsim \SI{12}{\micro\meter}$ is probably mesh-based.
This finding can not be automatically extrapolated to different pressure due to very different plasma densities.
For example, in our simulations with $n_\e\approx \SI{E18}{\per\cubic\meter}$, $\Delta x\approx \SI{10}{\micro\meter}$ and $N_\ppc = 64$, particles have an average weight $w \approx 15.6$ in the streamer head.
But in sprite discharges one may have $n_\e\approx \SI{E10}{\per\cubic\meter}$ and $\Delta x\approx \SI{1}{\meter}$, so $N_\ppc =64$ yields particle weights $w \approx \SI{1.56E8}{}$.
Discrete particle noise is therefore much higher for sprites than for atmospheric pressure streamers. 

\begin{figure}[h!t!b!]
  \centering
  \includegraphics{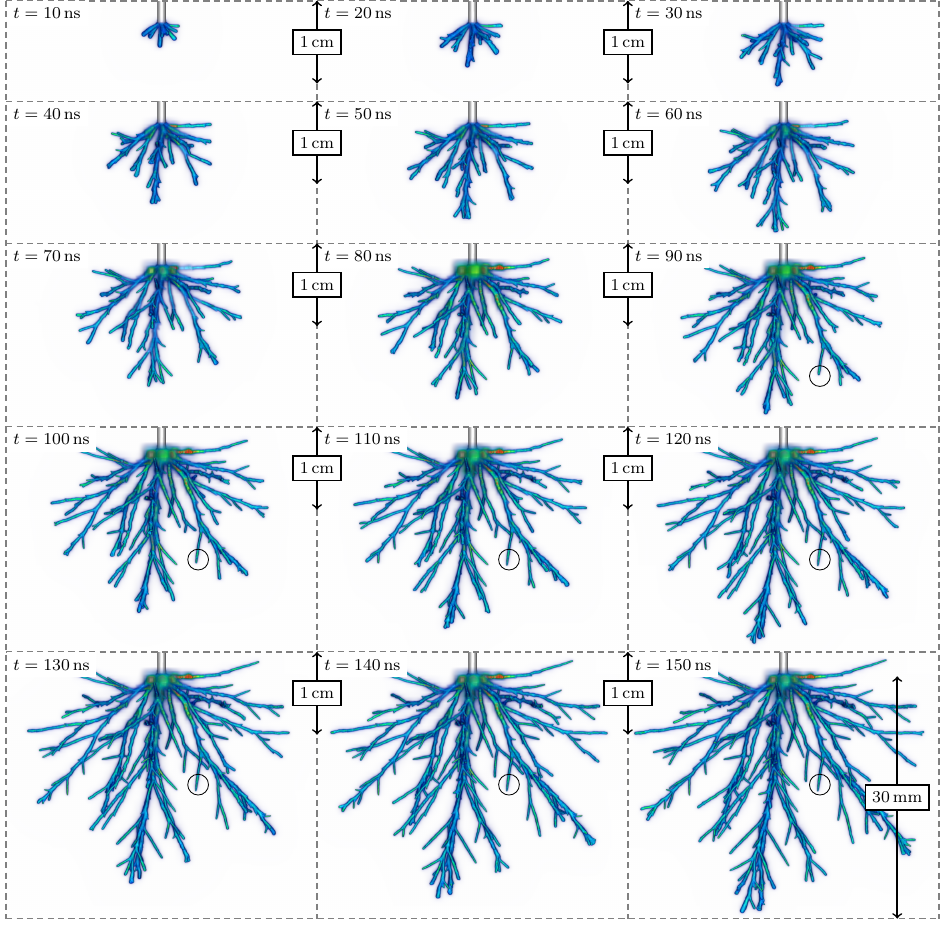}
  \caption{Snapshots of the electron density at various time instants.
    The circle shown in panels with $t\geq \SI{90}{\nano\second}$ indicates one of the stagnant streamer branches.
    }
  \label{fig:PositiveEvolution}
\end{figure}

\subsection{Positive streamer evolution}
\label{sec:positive_streamer}

We now run the simulation with $\Delta x\approx \SI{6}{\micro\meter}$ and $\Delta t=\SI{10}{\pico\second}$, using $N_\ppc=64$ particles per cell until $t=\SI{150}{\nano\second}$.
This simulation case is probably not completely grid converged, which we can see from the fact that the discharge tree in figure~\ref{fig:ItoConvergence} for $\Delta x=\SI{6}{\micro\meter}$ and $\Delta t=\SI{5}{\pico\second}$ is approximately \SI{25}{\percent} faster.
Part of this difference can also be due to a natural variation in the front velocity, depending on how the discharge tree develops.
Since we only ran a single simulation per spatial and temporal resolution, we do not know the size of these deviations.
Comparing with the panel $\Delta x=\SI{3}{\micro\meter}$ and $\Delta t=\SI{5}{\pico\second}$ in the same figure, we find that we are probably also missing some fine-scale features in the discharge tree as well.
Ideally, this is the simulation that we would have run further, but the case $\Delta x=\SI{6}{\micro\meter}$, $\Delta t=\SI{10}{\pico\second}$ was the largest case we could fit in our current compute quota. 

The simulation case is nonetheless quite challenging.
Positive streamers in air have smaller radii and branch more frequently than negative ones, and the branching behavior is also voltage-dependent as positive streamers in air branch more frequently at lower background fields.
In our case the average background electric field measured along the symmetry axis is just \SI{0.4}{\kilo\volt\per\milli\meter}, which is below the so-called positive streamer stability field of $\sim \SI{0.5}{\kilo\volt\per\milli\meter}$.
Under these conditions, experiments show that repeated branching leads to development of a discharge tree consisting of multiple small-diameter positive streamers \cite{Briels2008}.
Streamers may also stagnate, which has been identified as a challenge for fluid models \cite{Pancheshnyi2004,Marskar2020,Niknezhad2021,Li2022}.


Figure~\ref{fig:PositiveEvolution} shows temporal snapshots of the positive streamer evolution every ten nanoseconds until $t=\SI{150}{\nano\second}$.
We have not (yet) been able to skeletonize the discharge structure for quantitative analysis, but from the figure(s) we extract the following information:

\begin{itemize}
\item The streamer radii vary between \SI{\sim 1}{\milli\meter} and \SI{\sim 100}{\micro\meter}.
  The thicker streamers appear closest to the anode, and as they propagate they branch into thinner filaments.
  
\item In addition to propagating towards to the ground plane, electrostatic repulsion between the filaments yields numerous sideways branches, which determines the radius of the tree.
  
\item The front velocity is on average \SI{0.2}{\milli\meter\per\nano\second}, and is defined by the velocity of the front streamers.
  
\item The total length of the discharge tree at $t=\SI{150}{\nano\second}$ is approximately \SI{3}{\centi\meter} and its radius is approximately $\SI{1.5}{\centi\meter}$.
  
\item Many streamer filaments stop propagating after some distance, but do not lead to unbounded growth in the plasma density.
  We have indicated one of these branches in figure~\ref{fig:PositiveEvolution}, but many more can be identified.
\end{itemize}

\begin{figure}[h!t!b!]
  \centering
  \includegraphics{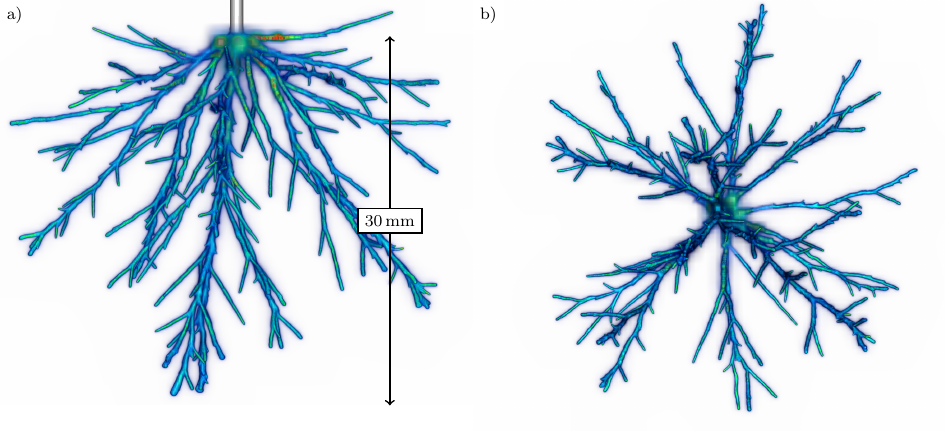}
  \caption{Different views of the positive streamer after $t=\SI{150}{\nano\second}$.
    a) Side view.
    b) Looking into the streamer from below.
  }
  \label{fig:PositiveViews}
\end{figure}

Figure~\ref{fig:PositiveViews} shows the discharge at $t=\SI{150}{\nano\second}$ from additional perspectives.
This data corresponds to the bottom-right panel in figure~\ref{fig:PositiveEvolution}.
As an amendment to figure~\ref{fig:PositiveEvolution} and figure~\ref{fig:PositiveViews}, we have added an animation of the corresponding data to the supplemental material in this article.

In the simulations we do not observe streamer merging, which can occur if two streamer heads are sufficiently close in space and there is sufficient photoionization between them \cite{Luque2008}.
Although streamer merging is fundamentally possible under very specific conditions, the prerequisites for it to occur are obviously not present in our simulations.
Streamer reconnection \cite{Nijdam2009} is not observed in the simulations either.
Reconnection occurs when a streamer connects into the wake of another streamer, and is thus a different phenomenon from streamer merging where two streamer heads merge directly.
We have observed streamer reconnection in other 3D fluid simulations when one of the streamer filaments reach the ground plane and acts as a virtual ground for the other streamers.
Further details regarding this finding will be reported elsewhere.

The simulation results presented here clearly can not be understood from axisymmetric simulations.
Many axisymmetric studies of streamer discharges have been reported in the last decades, but these only show the emergence of a single filament and therefore have a limited range of applicability.
The radius and velocity of the streamer then tend to increase with streamer length \cite{Bagheri2018} (depending on the field conditions), but experiments generally show repeated branching into thinner filaments \cite{Briels2008}.
Because thinner streamers propagate slower than thicker streamers, it is reasonable to expect that repeated branching lowers the front velocity.
It is obvious that the front velocity of the tree is determined by the velocity of the front streamers, but we point out that these streamers are also influenced by fields set up by neighboring branches.
The dynamics of such streamers and corresponding single-filament streamers with the same radii are therefore not necessarily the same.

In our simulations the streamers branch into small-diameter streamers, and the most relevant parametric 2D studies are the ones pertaining to so-called minimal streamers.
Such streamers have the smallest experimentally observed radius, do not branch, but still propagate over comparatively long distances.
In order to compare our front velocity with that of single streamers, we consider the case in \citet{Li2022} who presented a computational analysis of steady and stagnating positive streamers in air.
The reported velocities in \cite{Li2022} varied from \SIrange{0.25}{1.25}{\milli\meter\per\nano\second} when the streamer radii varied from \SIrange{25}{125}{\micro\meter}, and the corresponding electric field at the streamer tip varied from \SIrange{220}{150}{\kilo\volt\per\centi\meter}.
In our simulations we observe the same emerging radii and velocities, while the electric field is slightly higher at approximately \SIrange{\sim250}{\sim280}{\kilo\volt\per\centi\meter}.
We do not know the source of this discrepancy, but it could be due to the slightly larger numerical resolution used in our simulations.
Another factor could be that \citet{Li2022} use a correction to the $\alpha$-coefficient while we do not.
Both of these factors can facilitate small-scale features with higher fields.
Regardless of these finer points, the streamer radii, velocities, and fields that emerge in our simulations are consistent with the parametric study in \citet{Li2022}.

\subsection{Computational characteristics}
We now present some of the computational characteristics for the simulation in section~\ref{sec:positive_streamer}.
The simulation was run on 32 nodes on the Norwegian supercomputer Betzy, and ran to completion in about 4 days.
Each node on Betzy consists of two AMD Epyc 7742 CPUs for a total of 128 cores per node, so we used 4096 CPU cores in total. 
When we terminated the simulation it consisted of approximately 500 million grid cells and \SI{E10}{} computational particles, so there were approximately \num{15.6} million grid cells and \num{312.5} million particles per node.
Although our simulations are firmly footed in the realm of high-performance computing, they do not represent a large burden for modern supercomputers (which currently can have more than one million CPU cores).

\begin{figure}[h!t!b!]
  \centering
  \includegraphics{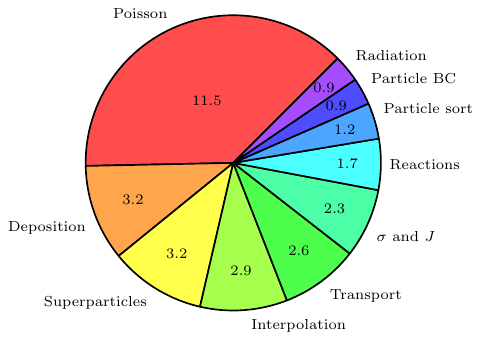}
  \caption{
    Computational characteristics, showing the time (in seconds) spent in various computational kernels.
  }
  \label{fig:CompLoad}
\end{figure}

Figure~\ref{fig:CompLoad} shows a breakdown of the kernel costs for the time step at $t = \SI{150}{\nano\second}$, showing the wall clock time spent in various computational routines.
This particular time step took around \SI{26}{\second}, but time steps varied down to around \SI{1}{\second} at the very beginning of the simulation.
Radiation transport for our simulations has a negligible cost, but this would not be the case e.g. for sprite simulations where many more physical photons are generated per cell.
The KMC algorithm has a cost of about \SIrange{5}{10}{\percent}, whereas setting up and solving the Poisson equation had a relative cost of approximately \SI{35}{\percent}.
Particle-related routines like superparticle handling, particle-mesh operations, transport, and spatial binning of particles had an accumulated cost of around 50\%.
There was also significant load imbalance for the super-particle handling since we load-balanced using the number of particles.
This distributed the load along the streamer channels, but did not account for particle merging/splitting which mainly occured in the streamer head.

Regridding the solution had about the same cost as 1-2 time steps, which was done every ten time steps.
The majority of the regrid cost comes from 1) recomputing cut-cell stencils and re-solving the Poisson equation on the new grids, and 2) redistributing particles on the new grids.
Particle redistribution is, unfortunately, quite expensive since virtually all computational particles change MPI rank ownership during regrids.

Two forms of I/O were used in the simulations: Plot files and checkpoint files.
Plot files contained data for analysis and ranged up to \SI{160}{\giga\byte} per file.
Checkpoint files contained data for restarting simulations, e.g. in case of hardware failures or for allocating more nodes as the simulation mesh grows.
These ranged up to \SI{360}{\giga\byte} per file.
Plot and checkpoint files took about two minutes to write, and were written every 100th time step.

\section{Conclusions and outlook}
\label{sec:conclusion}

\subsection{Main findings}

We have presented the foundation of a new type of model for streamer discharges based on a microscopic drift-diffusion model with a Kinetic Monte Carlo solver for the plasma chemistry.
A thematic discussion on the role of \ito-KMC and its connection to conventional fluid models was presented.
The model was coupled to photoionization with Monte-Carlo radiative transport, and a particle merging and splitting algorithm was presented. 
Suitable algorithms for integrating the equations of motion were then presented.
These algorithms were implemented in 2D and 3D and adapted to cut-cell Cartesian AMR grids.
We then implemented an example model for streamer discharges in air in needle-plane gaps, and showed that the \ito-KMC model agrees qualitatively and quantitatively with conventional fluid models.
Example simulations that demonstrate the output and stability of the \ito-KMC model were then provided.
The simulations presented in this paper demonstrate the feasibility of simulating discharge trees containing many streamer branches.

There are some advantages to using the new model, which are listed below:
\begin{itemize}
\item The model takes the same input as a fluid model, e.g. mobility and diffusion coefficients, and reaction rates.
\item It is inherently a PIC model and maintains particle discreteness.
\item \ito-KMC incorporates both reactive and diffusive fluctuations.
\item The model is exceptionally stable in both space and time, even on very coarse grids and for large time steps.
  The absence of a CFL condition is particularly liberating.  
\end{itemize}

We conjecture that the \ito-KMC method will be suitable for hybrid modeling \cite{Li2010,Li2012} where some of the electrons are treated kinetically.
Unlike hybrid models based on a fluid description, transfer of electrons between \ito-KMC and PIC-MCC descriptions can be done without disturbing the original charge distribution.

\subsection{Future work}

In this paper our focus has been on an all-discrete approach where also the ions are treated using \^Ito diffusion.
Future works will benefit from a mixed description where the electrons are treated using \^Ito diffusion while (some of) the heavy species are treated using a continuum model.
This can substantially reduce the computational load when more species of ions are tracked.
However, the relative cost of the Poisson solver is already quite high, and even with this improvement the Poisson equation will remain a computational bottleneck.

As the \ito-KMC is highly stable, higher order algorithms become particularly attractive.
In the absence of solid boundaries, fourth order deposition and interpolation methods exist \cite{Myers2017}, but these techniques have not been extended to cases where embedded boundaries are involved. 
For the Poisson equation, fourth order discretizations that include embedded boundary formulations have been reported \cite{Devendran2017}.
In time, only second-order convergence can be expected in the context of splitting methods and, furthermore, higher-order integration for particle transport is a significant challenge due to the presence of the Wiener process.
Suppressing numerical streamer branches for coarse-grid simulations is also desirable.
Moving forward, we will explore strategies for suppressing these by filtering the solutions \cite{Birdsall2004}.

The KMC-particle reactive coupling used in this paper adds some numerical diffusion into the system. 
Here, the algorithm is set up to uniformly distribute reactive products over a grid cell, and so it will end up placing secondary electrons in the wake of primary electrons.
This can become a source of numerical instability quite similar to the one seen in fluid simulations, but fortunately we have not (yet) observed these types of instabilities in our simulations.
Regardless, future efforts may benefit from reducing this source of numerical diffusion by introducing sub-grid models for the KMC-particle coupling.

We expect that 3D streamer simulations will become increasingly more sophisticated in the future.
In parallel with this development, there is an emergent need for improving the tools that we use for analyzing such simulations.
While two-dimensional simulations are relatively straightforward to analyze, 3D discharge trees are much more complex.
Quantitative analysis requires us to skeletonize the discharge trees in full 3D for extraction of branching ratios and angles, filament lengths, velocities, and so on.
In the future, we will also focus on establishing such analysis procedures.

\section*{Acknowledgements}
This study was partially supported by funding from the Research Council of Norway through grants 319930/E20 and 321449.
The computations were performed on resources provided by UNINETT Sigma2 - the National Infrastructure for High Performance Computing and Data Storage in Norway.
The author expresses his gratitude to Fanny Skirbekk for providing the images used in Fig.~\ref{fig:Nijdam}.

\section*{Code availability statement}
The computer code that was used to perform the calculations in this paper is publically available at \url{https://github.com/chombo-discharge/chombo-discharge}.
Input scripts are available upon reasonable request.

\bibliography{ltp, ebamr}

\end{document}